\documentclass[%
preprint,
 showpacs,preprintnumbers,
 nofootinbib,
 amsmath,amssymb,
 aps,
dvipdfmx
]{revtex4-1}

\usepackage{graphicx}

\usepackage{bm}
\usepackage{amsmath}
\usepackage{ulem}
\usepackage[usenames]{color}
\usepackage{comment}

\begin{document}



\newcommand{\bbra}[1]{\langle\!\langle {#1} |}     
\newcommand{\kket}[1]{| {#1} \rangle\!\rangle}     
\newcommand{\rbra}[1]{( {#1} |}     
\newcommand{\rket}[1]{| {#1} )}     
\newcommand{\rdket}[1]{|\!| {#1} )}     
\newcommand{\rrket}[1]{| {#1} ))}     
\newcommand{\dbra}[1]{\langle {#1} |\!|}     
\newcommand{\rdbra}[1]{( {#1} |\!|}     
\newcommand{\dket}[1]{|\!| {#1} \rangle}     
\newcommand{\mib}[1]{\boldsymbol {#1}}
\newcommand{\maru}[1]{\breve{#1}} 
\newcommand{\wtilde}[1]{\widetilde{#1}} 
\newcommand{\lsim}{{\stackrel{<}{\sim}}}
\newcommand{\gsim}{{\stackrel{\displaystyle >}{\raisebox{-1ex}{$\sim$}}}}
\newcommand{\lal}{\langle\!\langle}
\newcommand{\rar}{\rangle\!\rangle}
\newcommand{\wb}[1]{\overline{#1}}
\newcommand{\vect}[1]{\overrightarrow{#1}}
\newcommand{\ovl}[1]{\overline{#1}}
\newcommand{\braket}[1]{\langle{#1}\rangle}
\newcommand{\fsl}[1]{{\ooalign{\hfil/\hfil\crcr$#1$}}}
\def\beq{\begin{eqnarray}}
\def\eeq{\end{eqnarray}}
\def\bsub{\begin{subequations}}
\def\esub{\end{subequations}}
\def\beq{\begin{eqnarray}}
\def\eeq{\end{eqnarray}}
\def\bsub{\begin{subequations}}
\def\esub{\end{subequations}}
\def\b{\begin{equation}}
\def\bs{\begin{split}}
\def\es{\end{split}}
\def\e{\end{equation}}

\title{
A possibility of existence of a pseudovector-type quark-antiquark condensate in the quark matter and Nambu-Goldstone modes 
on that condensate in the Nambu-Jona-Lasinio model
}

\author{Kentaro Hayashi}
\affiliation{Graduate School of Integrated Arts and Science, Kochi University, Kochi 780-8520, Japan}

\author{Yasuhiko Tsue}
\email{tsue@kochi-u.ac.jp}
\affiliation{Department of Mathematics and Physics, Kochi University, Kochi 780-8520, Japan}




\begin{abstract}
A possibility of a pseudovector-type quark-antiquark condensed phase, 
which leads to a quark spin polarized phase, in the quark matter 
is investigated taking account of the vacuum effects 
leading to the chiral symmetry breaking by using the Nambu-Jona-Lasinio model. 
Also, possible Nambu-Goldstone modes on the pseudovector-type quark-antiquark condensate 
and the tensor-type quark-antiquark condensate, which also leads to the quark spin polarized phase, 
are investigated.
\end{abstract}



\maketitle

\section{Introduction}

One of recent interests with respect to the strong interaction governed by the quantum chromodynamics (QCD) 
may be to clarify the phase structure on the plane spanned by the temperature and density \cite{FH}. 
Especially, at finite density and low temperature, some authors have pointed out 
\cite{NMT,TMN,Maedan,Morimoto2}
that there exists a possibility of 
a quark spin polarization due to a pseudovector-type quark-antiquark condensation by using the low energy effective model 
of QCD such as the Nambu-Jona-Lasinio (NJL) model \cite{NJL}. 
Also, as was shown in the previous work, the spin polarization may also occur due to a tensor-type quark-antiquark condensate 
in the NJL model \cite{YT2015} and the magnetic moment density is evaluated in the hybrid star with a quark matter 
in the inner core of compact stars \cite{Matsuoka2018}.

By carrying out the Weinberg transformation, the pseudovector condensate due to the pseudovector 
interaction in the NJL model can be recast into the inhomogeneous condensate \cite{oursIJMP}. 
Under the inhomogeneous condensate, the excited modes, especially the Nambu-Goldstone (NG) modes 
\cite{NJL,G,GSW}
have been investigated by the Landau-Ginzburg-Wilson effective Lagrangian \cite{LNTTF,Kamikado}.

In this paper, a possibility of the existence of the pseudovector-type quark-antiquark condensed phase 
in the quark matter is investigated and its possibility is indicated taking account of the vacuum effect that 
leads to the chiral symmetry breaking, which may be neglected in some previous investigations \cite{Maedan, Morimoto2}. 
Further, the massless Nambu-Goldstone modes on the pseudovector condensate 
due to the psedovector interaction in the NJL model, which leads to the quark spin polarization, 
are investigated directly.

We consider two condensates. 
One is a usual chiral condensate $\braket{{\bar \psi}\psi}$. 
Another is the pseudovector condensate, namely 
$\braket{{\bar \psi}\gamma_5\gamma^{\mu=3}\tau_{i=3}\psi}=\braket{\psi^{\dagger}\Sigma_3\tau_3\psi}$, 
where $\Sigma_3$ represents the third component of the spin matrix and 
$\gamma^\mu$ and $\tau_i$ represent the Dirac gamma matrices and the isospin matrices, respectively. 
Although the quark spin polarization, of course, does not appear in vacuum, 
the quark spin polarization may occur in a certain region of the QCD phase diagram 
such as the high density and low temperature quark matter. 
It will be shown that the pseudovector condensed phase may appear in a certain region with respect to 
the quark chemical potential. 
Therefore, in consideration of a possibility of existence of both the pseudovector condensate and the chiral condensate 
in a certain parameter region, the NG modes will be investigated in the case of the coexistence of the psuedovector condensate 
and the chiral condensates as well as in the case of the pseudovector condensate only. 
In addition to the above two condensates, we consider the tensor condensate, namely 
$\braket{{\bar \psi}\gamma^{\mu=1}\gamma^{\mu=2}\tau_{i=3}\psi}=\braket{{\bar \psi}\Sigma_3\tau_3\psi}$. 
This condensate does not coexist with the chiral condensate as was shown in the previous paper \cite{Matsuoka2016}. 
Thus, the NG modes will be also investigated on the tensor condensate only.

This paper is organized as follows: 
In the next section, the gap equations for both the chiral condensate and the pseudovector condensate 
are derived in the NJL model with 
the pseudovector interaction between quarks. 
In section 3, the numerical results for the gap equations are shown and 
the possibility of the existence of the psuedovector condensed phase at finite density is indicated. 
In section 4, the massless modes are investigated on the pseudovector condensate.  
Also, in section 5, the gap equations for the tensor-type quark-antiquark condensate, which may also leads to 
the quark spin polarization, is revisited following to the previous paper \cite{Matsuoka2016} 
in the NJL model with a tensor interaction between quarks. 
The detailed calculations to derive the massless modes are shown in Appendix A. 
The last section is devoted to a summary and concluding remarks.

\section{Gap equations with respect to the chiral condensate and the pseudovector condensate}

Let us introduce the NJL model with vector and pseudovector interactions which retains the chiral 
symmetry \cite{NJL,Klevansky,HK,Buballa}: 
\begin{equation}\label{2-1}
 \mathcal{L}_{\rm NJL}=\bar{\psi}i\gamma^\mu\partial_\mu\psi+G_S[(\bar{\psi}\psi)^2+(\bar{\psi}i\gamma_5\bm{\tau}\psi)^2]
  +G[(\bar{\psi} \gamma^\mu\bm{\tau}\psi)^2+(\bar{\psi}\gamma_5\gamma^\mu\bm{\tau}\psi)^2] \ ,
\end{equation}
where $\bm{\tau}$ represents the isospin matrices with respect to the flavor $su(2)$-symmetry.
The generating functional $Z$ is written as 
\beq\label{2-2}
Z=\int{\cal D}\psi{\cal D}{\bar{\psi}}\ \exp \left(i\int d^4x {\cal L}_{\rm NJL}\right)\ .
\eeq
Here, we use the auxiliary field method and introduce the functional Gaussian integral leading to unit as 
\beq\label{2-3}
& &1=\int{\cal D}\sigma'{\cal D}{\bm{\pi}}'\ \exp\left(i\int d^4x \left[-\frac{1}{G_S}
\left(\sigma'{}^2+\bm{\pi}'{}^2\right)\right]\right)\ , \nonumber\\
& &1=\int{\cal D}\bm{\rho}'{}^\mu{\cal D}{\bm{a}}'{}^\mu\ \exp\left(i\int d^4x \left[-\frac{1}{G}
\left(\bm{\rho}'{}^\mu\cdot{\bm{\rho}}'_\mu+\bm{a}'{}^\mu\cdot\bm{a}'_\mu\right)\right]\right)\ . 
\eeq
Inserting the above unit into the generating functional $Z$ in Eq.(\ref{2-2}) and 
replacing $\sigma'$, $\bm{\pi}'$, $\bm{\rho}'{}^\mu$ and $\bm{a}'{}^\mu$ into 
$\sigma$, $\bm{\pi}$, $\bm{\rho}^\mu$ and $\bm{a}^\mu$ as 
\beq\label{2-4}
& &\sigma'=\sigma +G_S(\bar{\psi}\psi) \ , \nonumber\\
& &\bm{\pi}' = \bm{\pi}+G_S(\bar{\psi}i\gamma_5\bm{\tau}\psi) \ , \nonumber\\
& &\bm{\rho}'{}^\mu=\bm{\rho}^\mu+G(\bar{\psi}\gamma^\mu\bm{\tau}\psi) \ , \nonumber\\ 
& &\bm{a}'{}^\mu =\bm{a}^\mu+G(\bar{\psi}\gamma_5\gamma^\mu\bm{\tau}\psi) \ ,
\eeq
the generating functional is rewritten as 
\beq\label{2-5}
Z&=&\int{\cal D}\psi{\cal D}{\bar{\psi}}\int{\cal D}\sigma{\cal D}{\bm{\pi}}{\cal D}\bm{\rho}^\mu{\cal D}{\bm{a}}^\mu
\exp \left(i\int d^4x {\wtilde{\cal L}}\right)\ ,
\eeq
where 
\beq\label{2-6}
{\wtilde {\cal L}}&=&
\bar{\psi}(i\gamma^\mu\partial_\mu - 2\sigma - 2\bm{\pi}\cdot i\gamma_5\bm{\tau} 
                     - 2\bm{\rho}_\mu\cdot\gamma^\mu\bm{\tau} - 2\bm{a}_\mu\cdot\gamma_5\gamma^\mu\bm{\tau})\psi 
\nonumber\\
& & - \frac{1}{G_s}(\sigma^2+\bm{\pi}^2) 
                     - \frac{1}{G}(\bm{\rho}^\mu\cdot\bm{\rho}_\mu+\bm{a}^\mu\cdot\bm{a}_\mu)\ .
\eeq
Here, from $\delta {\wtilde {\cal L}}/\delta \sigma=0$ and so on, we obtain 
\beq\label{2-7}
& &\sigma =-G_S(\bar{\psi}\psi) \ , \nonumber\\
& &\bm{\pi}=-G_S(\bar{\psi}i\gamma_5\bm{\tau}\psi) \ , \nonumber\\
& &\bm{\rho}_\mu=-G(\bar{\psi}\gamma_\mu\bm{\tau}\psi) \ , \nonumber\\ 
& &\bm{a}_\mu=-G(\bar{\psi}\gamma_5\gamma_\mu\bm{\tau}\psi) \ .
\eeq
Further, integrating $\psi$ and ${\bar \psi}$, the generating functional can be expressed as 
\beq\label{2-8}
Z&=&\int{\cal D}\sigma{\cal D}{\bm{\pi}}{\cal D}\bm{\rho}^\mu{\cal D}{\bm{a}}^\mu
\exp \left(i\Gamma\right)\ ,\nonumber\\
\Gamma&=&\int d^4 x\left[-\frac{1}{G_s}(\sigma^2+\bm{\pi}^2) 
                     - \frac{1}{G}(\bm{\rho}^\mu\cdot\bm{\rho}_\mu+\bm{a}^\mu\cdot\bm{a}_\mu)\right]
\nonumber\\
& &\qquad\quad 
-i\ln {\rm Det} \left(i\gamma^\mu\partial_\mu - 2\sigma - 2\bm{\pi}\cdot i\gamma_5\bm{\tau} 
                     - 2\bm{\rho}_\mu\cdot\gamma^\mu\bm{\tau} - 2\bm{a}_\mu\cdot\gamma_5\gamma^\mu\bm{\tau}\right)
\ . 
\eeq
It should be noted that the pseudovector mode $\bm{a}_3$ is written as 
\beq\label{2-9} 
\bm{a}_3&=&-G\ \bar{\psi}\gamma_5\gamma_3\bm{\tau}\psi \nonumber\\
&=&-G\ \psi^{\dagger}\Sigma_3 \bm{\tau} \psi\ .
\eeq
Here, we use the Dirac representation for the Dirac gamma matrices, namely, 
\beq\label{2-10}
& &\gamma^\mu=(\gamma^0, \bm{\gamma})\ , \qquad
\gamma^0=\begin{pmatrix}
		1 & 0  \\
		 0 & -1 
	\end{pmatrix} \ , \qquad
\bm{\gamma}=\begin{pmatrix}
		0 & \bm{\sigma}  \\
            -\bm{\sigma} & 0 
	\end{pmatrix} \ , \qquad
\gamma_5=\begin{pmatrix}
		0 & 1  \\
		1 & 0 
	\end{pmatrix} \ , \nonumber\\
& &\gamma^0\gamma_5\gamma_3\equiv\Sigma_3=\begin{pmatrix}
		\sigma_3 & 0  \\
		 0 & \sigma_3 
	\end{pmatrix} \ , 
\eeq
where $\bm{\sigma}$ and $\sigma_3$ represent the Pauli spin matrices and its third component. 
Thus, $\Sigma_3$ is the third component of the spin matrices.

Hereafter, let us assume that there exist the chiral condensate $\sigma_0=\braket{\sigma}$ 
and pseudovector condensate $a_0=\braket{a_{\mu=3}^{i=3}}$ 
with $\mu=3$ and the third component of the isospin, 
namely 
\beq\label{2-11}
& &\sigma_0\equiv \braket{\sigma}=-G_S\braket{{\bar \psi}\psi}\ , \nonumber\\
& &a_0\equiv \braket{a_{\mu=3}^{i=3}}=-G\braket{\psi^{\dagger}\Sigma_3\tau_3 \psi}\ . 
\eeq
Thus, the pseudovector condensate $\braket{a_{\mu=3}^{i=3}}$ can be regarded as the 
quark spin polarization. 
Under these quark-antiquark condensates, the effective potential $V(\sigma_0,a_0)$ can be 
derived from the effective action $\Gamma$ as
\beq\label{2-12}
\Gamma(\sigma=\sigma_0,a_3^3=a_0)&=&-V(\sigma_0,a_0)\int d^4 x\ , \nonumber\\
V(\sigma_0,a_0)
&=&\frac{1}{G_S}\sigma_0^2-\frac{1}{G}a_0^2-
\int\frac{d^4 k}{i(2\pi)^4}\ln\det \left[
\gamma^\mu k_\mu -2\sigma_0-2a_0\gamma_5\gamma^3\tau_3\right]\nonumber\\
&=&\frac{1}{G_S}\sigma_0^2-\frac{1}{G}a_0^2-
\int\frac{d^4 k}{i(2\pi)^4}{\rm tr}\ln \left[
\gamma^\mu k_\mu -2\sigma_0-2a_0\gamma_5\gamma^3\tau_3\right]\ , \qquad  
\eeq
where trace is taken by the Dirac gamma matrices, isospin space and color space. 
Here, it should be noted that $a^{\mu=3}_{i=3}a_{\mu=3}^{i=3}=-a_0^2$ is used because, in this paper, 
the metric tensor is adopted as 
\beq\label{2-13}
g^{\mu\nu}={\rm diag}.(1,\ -1,\ -1,\ -1)\ . 
\eeq

The condensates $\sigma_0$ and $a_0$ are determined by the following gap equations : 
\beq\label{2-14}
& &\frac{\partial V}{\partial \sigma_0}
=\frac{2}{G_S}\sigma_0
-\int \frac{d^4 k}{i(2\pi)^4}
{\rm tr} \left[\frac{-2}{\gamma^\mu k_\mu -2\sigma_0-2a_0\gamma_5\gamma^3\tau_3}\right]=0\ , \nonumber\\
& &\frac{\partial V}{\partial a_0}
=-\frac{2}{G}a_0
-\int \frac{d^4 k}{i(2\pi)^4}
{\rm tr} \left[\frac{1}{\gamma^\mu k_\mu -2\sigma_0-2a_0\gamma_5\gamma^3\tau_3}(-2\gamma_5\gamma^3\tau_3)\right]=0\ , 
\eeq 
which, after some tedious calculation for the gamma matrices, lead to 
\beq\label{2-15}
& &\sigma_0\left[
1+48G_S\int \frac{d^4 k}{i(2\pi)^4}
\frac{k^2-4\sigma_0^2+4a_0^2}{(k^2-4\sigma_0^2-4a_0^2)^2-16a_0^2(k_3^2+4\sigma_0^2)}\right]=0\nonumber\\
& &a_0\left[
1-48G\int \frac{d^4 k}{i(2\pi)^4}
\frac{k^2+4\sigma_0^2-4a_0^2+2k_3^2}{(k^2-4\sigma_0^2-4a_0^2)^2-16a_0^2(k_3^2+4\sigma_0^2)}\right]=0\ , 
\eeq
where $k^2=k_0^2-\bm{k}^2$ and $k_3$ means the third component of the space vector $\bm{k}$.

\section{Numerical evaluations for the solutions of the gap equations}

The NJL model is a cutoff theory. 
We determine the three-momentum cutoff $\Lambda$ and the coupling constant $G_S$ in order to 
reproduce the pion decay constant $f_{\pi}=92$ MeV and the quark mass 
$M=2\sigma_0=313$MeV. 
In this paper, we take $\Lambda=643$ MeV and $G_S=5.20$ GeV${}^{-2}$. 
However, the interaction strength $G$ for the pseudovector interaction cannot be determined. 
We here take $G$ as a free parameter. 

In this section only, the notations are changed for simplicity as follows: 
\beq\label{3-1}
G=-G_{pv}\ , \qquad \sigma_0=\frac{\sigma_s}{2}\ , \qquad a_0=\frac{\sigma_{pv}}{2}\ . 
\eeq
Of course, $\sigma_s$ corresponds to the quark mass $M$. 
Further, in order to consider the quark matter with finite quark chemical potential $\mu$, we add one term 
$\mu\psi^\dagger \psi$ in the effective Lagrangian density (\ref{2-6}). 
Thus, we should replace $k_0$ into $k_0+\mu$ in the gap equations. 
Then, the gap equations (\ref{2-15}) are recast into 
\beq\label{3-2}
& &\sigma_s\left[
1+48G_S\int \frac{d^4 k}{i(2\pi)^4}
\frac{(k_0+\mu)^2-{\mib k}^2-\sigma_s^2+\sigma_{pv}^2}
{((k_0+\mu)^2-{\mib k}^2-\sigma_s^2-\sigma_{pv}^2)^2-4\sigma_{pv}^2(k_3^2+\sigma_s^2)}\right]=0\nonumber\\
& &\sigma_{pv}\left[
1+48G_{pv}\int \frac{d^4 k}{i(2\pi)^4}
\frac{(k_0+\mu)^2-{\mib k}^2+\sigma_s^2-\sigma_{pv}^2+2k_3^2}
{((k_0+\mu)^2-{\mib k}^2-\sigma_s^2-\sigma_{pv}^2)^2-4\sigma_{pv}^2(k_3^2+\sigma_s^2)}\right]=0\ .
\nonumber\\
& &
\eeq

If there exist the nontrivial solutions $\sigma_s\neq 0$ and/or $\sigma_{pv}\neq 0$ of the above gap equations, 
we can rewrite the gap equations introducing the three-momentum cutoff $\Lambda$ as 
\beq
\sigma_s&=&
-\frac{24}{(2\pi)^2}G_S\int_0^\Lambda dk_{12}\int_0^{\sqrt{\Lambda^2-k_{12}^2}}dk_3 \nonumber\\
& &\qquad\qquad\qquad\qquad \times k_{12}
\biggl[\frac{\sigma_s-\frac{2\sigma_s\sigma_{pv}}{\sqrt{k_3^2+\sigma_s^2}}}{\epsilon_{\mib k}^{(-)}}
\left(\theta(\mu-\epsilon_{\mib k}^{(-)})-\theta(\mu+\epsilon_{\mib k}^{(-)})\right)
\nonumber\\
& &\qquad\qquad\qquad\qquad\qquad
+\frac{\sigma_s+\frac{2\sigma_s\sigma_{pv}}{\sqrt{k_3^2+\sigma_s^2}}}{\epsilon_{\mib k}^{(+)}}
\left(\theta(\mu-\epsilon_{\mib k}^{(+)})-\theta(\mu+\epsilon_{\mib k}^{(+)})\right)
\biggl]\ , 
\label{3-3}\\
\sigma_{pv}&=&
-\frac{24}{(2\pi)^2}G_{pv}\int_0^\Lambda dk_{12}\int_0^{\sqrt{\Lambda^2-k_{12}^2}}dk_3 \nonumber\\
& &\qquad\qquad\qquad\qquad \times
k_{12}
\biggl[\frac{\sigma_{pv}-\sqrt{k_3^2+\sigma_s^2}}{\epsilon_{\mib k}^{(-)}}
\left(\theta(\mu-\epsilon_{\mib k}^{(-)})-\theta(\mu+\epsilon_{\mib k}^{(-)})\right)\nonumber\\
& &\qquad\qquad\qquad\qquad\qquad
+\frac{\sigma_{pv}+\sqrt{k_3^2+\sigma_s^2}}{\epsilon_{\mib k}^{(+)}}
\left(\theta(\mu-\epsilon_{\mib k}^{(+)})-\theta(\mu+\epsilon_{\mib k}^{(+)})\right)
\biggl]\ , \ \ \ \ 
\label{3-4}
\eeq
where 
\beq\label{3-5}
\epsilon_{\mib k}^{(\pm)}=\sqrt{{\mib k}^2+\sigma_s^2+\sigma_{pv}^2\pm 2\sigma_{pv}\sqrt{k_3^2+\sigma_s^2}}\ , 
\qquad 
k_{12}=\sqrt{k_1^2+k_2^2}\ . 
\eeq
In Eqs. (\ref{3-3}) and (\ref{3-4}), 
the terms with $\theta(\mu+\epsilon_{\mib k}^{(\pm)})$ in the right-hand sides 
represent the 
vacuum effects which should not be omitted in this paper, while these effects were often neglected in the previous work.

\begin{figure}[t]
 \begin{center}
 \includegraphics[scale=0.55]{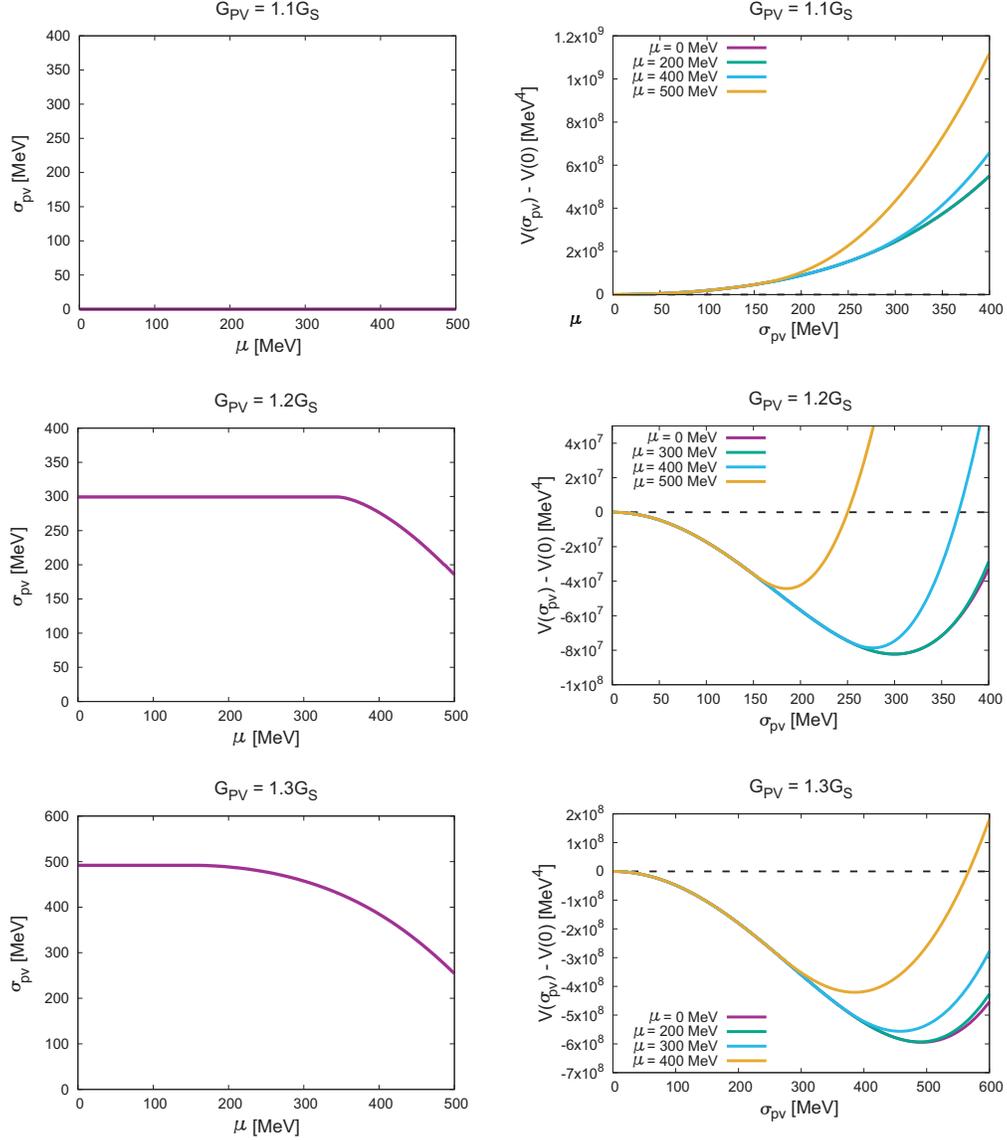}
 \caption{The effective potentials (right panels) and the pseudovector 
condensates $\sigma_{pv}$ (left panels) are shown with the pseudovector interaction strength 
$G_{pv}$=1.1$G_S$, $1.2G_S$ and $1.3G_S$, respectively. }
 \label{fig:fig1}
 \end{center}
\end{figure}

\begin{figure}[b]
 \begin{center}
 \includegraphics[scale=0.6]{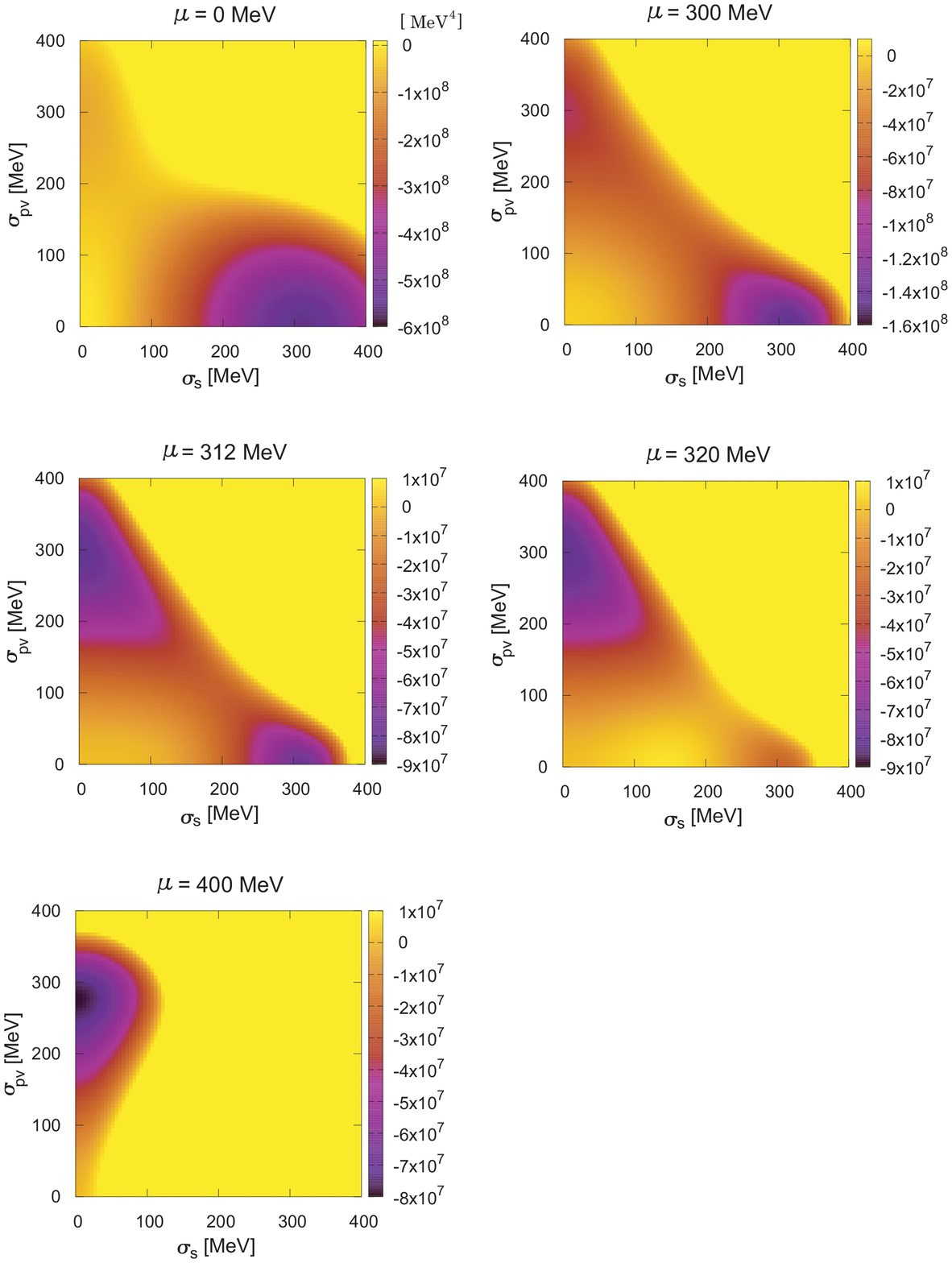}
 \caption{The contour plots of the effective potential as a function of the chiral and 
the psuedovector condensates, $\sigma_s$ and $\sigma_{pv}$, are depicted with the 
quark chemical potential $\mu=0$, 300 MeV, 312 MeV, 320 MeV and 400 MeV, respectively, 
in the case of $G_{pv}=1.2 G_S$. }
 \label{fig:fig2}
 \end{center}
\end{figure}

Similarly, the effective potential (\ref{2-12}) is calculated as follows:
\beq\label{6-6}
V(\sigma_s,\sigma_{pv})&=&
\frac{\sigma_s^2}{4G_S}+\frac{\sigma_{pv}^2}{4G_{pv}}\nonumber\\
& &-\frac{12}{(2\pi)^2}\int_0^\Lambda dk_{12}\int_0^{\sqrt{\Lambda^2-k_{12}^2}}dk_3
k_{12}[(\epsilon_{\mib k}^{(-)}-\mu)\theta(\epsilon_{\mib k}^{(-)}-\mu)
\nonumber\\
& &\qquad\qquad\qquad\qquad\qquad\qquad\qquad\qquad
+(\epsilon_{\mib k}^{(+)}-\mu)\theta(\epsilon_{\mib k}^{(+)}-\mu)] \ . 
\eeq

Let us, first, consider the effective potential without the chiral condensate, namely $\sigma_s=0$, 
in order to show the existence of the nontrivial solution $\sigma_{pv}\neq 0$ 
at finite quark chemical potential. 
Figure \ref{fig:fig1} shows the effective potential (right panels) and the pseudovector 
condensate $\sigma_{pv}$ (left panels) at some values of the pseudovector interaction strength 
$G_{pv}$, respectively. 
In Fig.\ref{fig:fig1}, the effective potential $V(\sigma_{pv})$ is normalized by $V(\sigma_{pv}=0)=0$, 
namely $V(\sigma_{pv})=V(\sigma_s=0, \sigma_{pv})-V(\sigma_s=0, \sigma_{pv}=0)$. 
It is shown that, in the case of $G_{pv}=1.1G_S$, the effective potential has a minimum at $\sigma_{pv}=0$ 
in all the chemical potential $\mu$. 
However, in the case above $G_{pv}=1.2G_S$, the effective potential has a minimum 
at a point with $\sigma_{pv}\neq 0$. 
Of course, as $G_{pv}$ becomes larger, then $\sigma_{pv}$ becomes larger. 
Thus, there is a possibility that the chiral symmetry is broken without the chiral condensate, 
$\sigma_s=0$, for rather large pseudovector interaction strength. .

\begin{figure}[b]
 \begin{center}
 \includegraphics[scale=0.6]{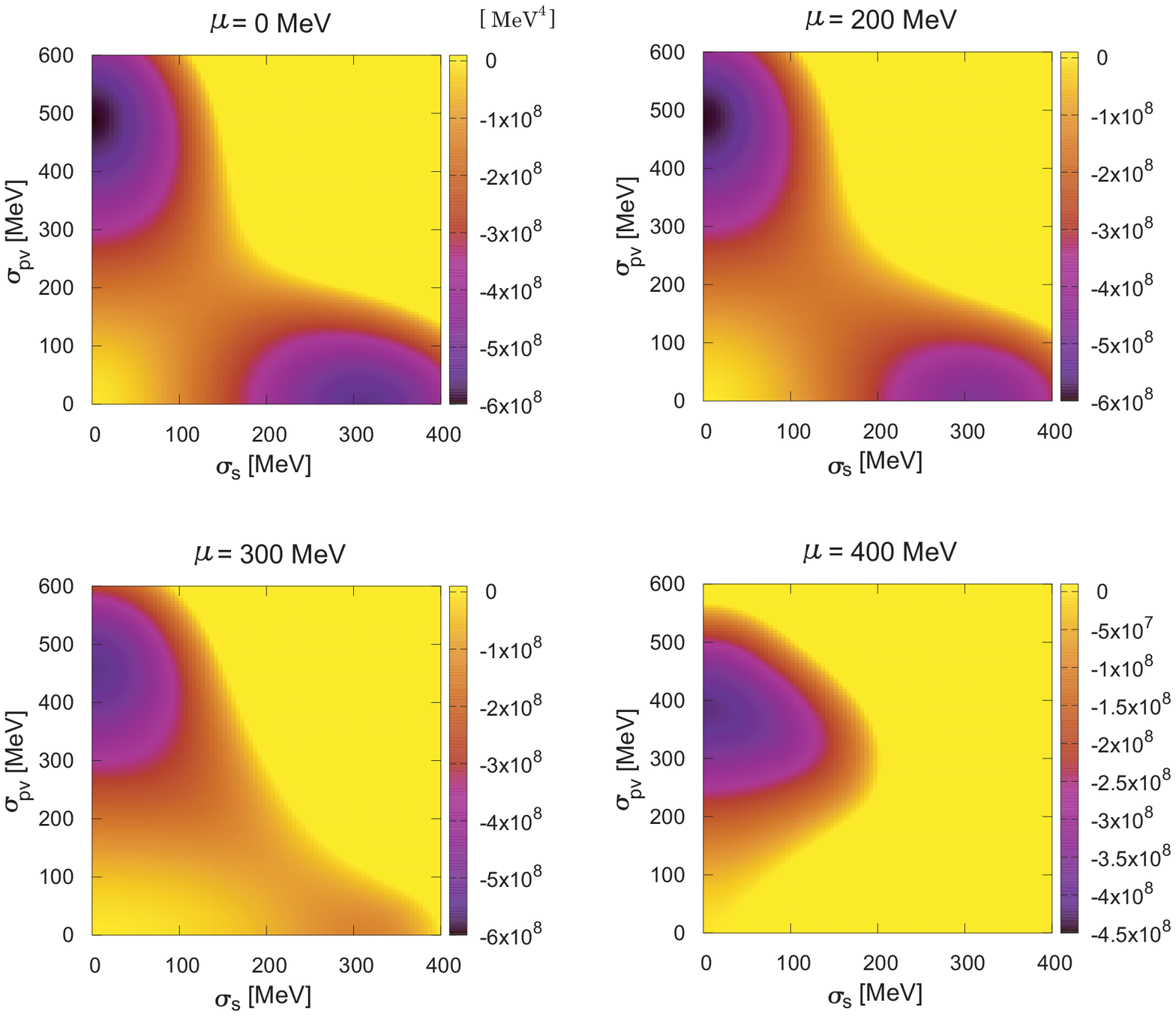}
 \caption{The contour plots of the effective potential as a function of the chiral and 
the psuedovector condensates, $\sigma_s$ and $\sigma_{pv}$, are depicted with the 
quark chemical potential $\mu=0$, 200 MeV, 300 MeV and 400 MeV, respectively, 
in the case of $G_{pv}=1.3 G_S$.}
 \label{fig:fig3}
 \end{center}
\end{figure}

Next, let us consider both the condensates $\sigma_s$ and $\sigma_{pv}$ simultaneously. 
In Fig.{\ref{fig:fig2}}, the contour plots of the effective potential as a function of the chiral and 
the psuedovector condensates, $\sigma_s$ and $\sigma_{pv}$, are depicted with different values of the 
quark chemical potential $\mu$ in the case of $G_{pv}=1.2 G_S$. 
At the quark chemical potential being 0, the absolute minimum point of the effective potential is given 
by the nonzero chiral condensate $\sigma_s$ and $\sigma_{pv}=0$. 
As the quark chemical potential becomes larger, the local minimum of the effective potential 
develops at $\sigma_{pv}\neq 0$ and $\sigma_s=0$. 
However, the absolute minimum is still the point with $\sigma_s \neq 0$ and $\sigma_{pv}=0$. 
On the other hand, around $\mu=312$ MeV, 
the local minimum with $\sigma_s=0$ and $\sigma_{pv}\neq 0$ is converted to the absolute minimum 
and vice versa. 
In the region with larger quark chemical potential, such as $\mu \approx 400$ MeV, 
the local minimum with $\sigma_s \neq 0$ disappears. 
Thus, the first order phase transition is seen from the chiral condensed phase to the 
pseudovector condensed phase.

Figure \ref{fig:fig3} is the same as Fig.\ref{fig:fig2} except for the pseudovector 
interaction strength, namely $G_{pv}=1.3G_S$ instead of $G_{pv}=1.2G_S$.  
In this strong coupling case, the absolute minimum of the effective potential gives 
nonzero $\sigma_{pv}$ and $\sigma_s=0$ even at the quark chemical potential $\mu$ being 0. 
On the other hand, the usual vacuum with $\sigma_s\neq 0$ and $\sigma_{pv}=0$ corresponds to 
the local minimum of the effective potential. 
This situation is unphysical because, if this situation is realized, 
it results in the quark spin polarization due to the pseudovector condensate in vacuum. 
Thus, even if there is the pseudovector interaction between quarks, the interaction strength 
must be restricted.

\section{Massless meson modes}

In this section, the massless Nambu-Goldstone modes on the chiral and/or pseudovector condensates are investigated.  
The meson propagator $\Delta_\alpha(x,y)$ can be derived from the effective action $\Gamma$ as 
\beq\label{3-16}
\Delta_{\alpha}^{-1}(x,y)=-\frac{\delta^2 \Gamma}{\delta \alpha (x) \delta \alpha(y)}\ , 
\eeq
where $\alpha(x)=\sigma(x),\ \bm{\pi}(x),\ \bm{\rho}_{\mu}(x)$ or $\bm{a}_{\mu}(x)$. 
In the momentum representation, the two-point vertex function $\Gamma_{\alpha}(p)$ can be expressed as 
\beq\label{3-17}
\Gamma_{\alpha}(p)(2\pi)^4\delta^4(p+q)
=\left. \frac{\delta^2 \Gamma}{\delta \alpha (p) \delta \alpha(q)}\right|_{\sigma=\sigma_0,a_3^3=a_0,\bm{\pi}=\bm{\rho}_{\mu}
=\bm{a}_{\mu\neq 3}=a_{\mu=3}^{i\neq 3}=0}\ ,
\eeq
where $p$ and $q$ represent the four-momenta. 
Therefore, for small $p^2$, $\Gamma_{\alpha}(p)$ behaves like 
\beq\label{3-18}
\Gamma_{\alpha}(p)=-\Delta_{\alpha}^{-1}(p)\propto m_\alpha^2-p^2+\cdots \ , 
\eeq
where $m_\alpha$ means the meson mass.

Going back to \S 2, two-point vertex functions are obtained as 
\beq
& &\Gamma_{\sigma}(p)=-\frac{2}{G_S}-\int\!\!\frac{d^4 k}{i(2\pi)^4}
{\rm tr}\left[
(-2)\frac{1}{\fsl{k}-2\sigma_0-2a_0\gamma_5\gamma^3\tau_3}\right.\nonumber\\
& &\qquad\qquad\qquad\qquad\qquad\qquad\quad
\times \left. (-2)
\frac{1}{\fsl{k}+\fsl{p}-2\sigma_0-2a_0\gamma_5\gamma^3\tau_3}\right]\ , 
\label{3-19}\\
& &\Gamma_{\pi,i}(p)=-\frac{2}{G_S}-\int\!\!\frac{d^4 k}{i(2\pi)^4}
{\rm tr}\left[
(-2i\gamma_5\tau_i)\frac{1}{\fsl{k}-2\sigma_0-2a_0\gamma_5\gamma^3\tau_3}\right. \nonumber\\
& &\qquad\qquad\qquad\qquad\qquad\qquad\qquad
\left. \times (-2i\gamma_5\tau_i)
\frac{1}{\fsl{k}+\fsl{p}-2\sigma_0-2a_0\gamma_5\gamma^3\tau_3}\right]\ , 
\label{3-20}\\
& &\Gamma_{\rho,i}^{\mu\nu}(p)=-\frac{2}{G}g^{\mu\nu}-\int\!\!\frac{d^4 k}{i(2\pi)^4}
{\rm tr}\left[
(-2\gamma^{\mu}\tau_i)\frac{1}{\fsl{k}-2\sigma_0-2a_0\gamma_5\gamma^3\tau_3}\right. \nonumber\\
& &\qquad\qquad\qquad\qquad\qquad\qquad\qquad\ \ 
\left. \times (-2\gamma^\nu\tau_i)
\frac{1}{\fsl{k}+\fsl{p}-2\sigma_0-2a_0\gamma_5\gamma^3\tau_3}\right]\ , 
\label{3-21}\\
& &\Gamma_{a,i}^{\mu\nu}(p)=-\frac{2}{G}g^{\mu\nu}-\int\!\!\frac{d^4 k}{i(2\pi)^4}
{\rm tr}\left[
(-2\gamma_5\gamma^{\mu}\tau_i)\frac{1}{\fsl{k}-2\sigma_0-2a_0\gamma_5\gamma^3\tau_3} \right. \nonumber\\
& &\qquad\qquad\qquad\qquad\qquad\qquad\qquad\ \ 
\left. \times (-2\gamma_5\gamma^\nu\tau_i)
\frac{1}{\fsl{k}+\fsl{p}-2\sigma_0-2a_0\gamma_5\gamma^3\tau_3}\right]\ , \quad
\label{3-22}
\eeq
where $\fsl{k}=\gamma^\mu k_\mu$ and so on.

\subsection{The case of $\sigma_s(=2\sigma_0) \neq 0$ and $\sigma_{pv}(=2a_0)=0$}

The chiral symmetry breaking usually occurs due to the chiral condensate $\braket{{\bar \psi}\psi}=-\sigma_0/G_S$. 
In this subsection, we treat the case of $\sigma_0(=\sigma_s/2)\neq 0$ and $a_0(=\sigma_{pv}/2)=0$. 
This case is nothing but the dynamical chiral symmetry breaking in the original NJL model. 
Therefore, as is well known, the NG boson is pion. 
Here, for the caution's sake, let us review its treatment. 
From Eq.(\ref{2-15}), the gap equation for $\sigma_0\neq 0$ and $a_0=0$ is as follows: 
\beq\label{3-23}
1+48G_S\int \frac{d^4 k}{i(2\pi)^4}
\frac{1}{k^2-4\sigma_0^2}=0 \ .
\eeq
Then, the inverse propagator in Eq.(\ref{3-20}) can be calculated as 
\beq\label{3-24}
\Gamma_{\pi,i}(p\rightarrow 0)
&=&-\frac{2}{G_S}-\int\!\!\frac{d^4 k}{i(2\pi)^4}
{\rm tr}\left[
(-2i\gamma_5\tau_i)\frac{1}{\fsl{k}-2\sigma_0} (-2i\gamma_5\tau_i)
\frac{1}{\fsl{k}-2\sigma_0}\right] \nonumber\\
&=&-\frac{2}{G_S}+4\int\!\!\frac{d^4 k}{i(2\pi)^4}
{\rm tr}\left[\frac{1}{-\fsl{k}-2\sigma_0}\cdot\frac{1}{\fsl{k}-2\sigma_0}\right]\nonumber\\
&=&-\frac{2}{G_S}-96\int\!\!\frac{d^4 k}{i(2\pi)^4}
\frac{1}{k^2-4\sigma_0^2}\nonumber\\
&=&0\ , 
\eeq
where the gap equation (\ref{3-23}) is used in the last line in (\ref{3-24}). 
Thus, from Eq.(\ref{3-18}) with $p\rightarrow 0$, all pion masses with any isospin component $i$ are equal to zero. 
Therefore, the pions are the NG bosons.

\subsection{The case of $\sigma_s(=2\sigma_0) =0$ and $\sigma_{pv}(=2a_0)\neq 0$}

Next, let us consider the case with $\sigma_0=0$ and $a_0\neq 0$. 
In this case, the gap equation in Eq.(\ref{2-15}) can be expressed as 
\beq\label{3-25}
1-48G\int \frac{d^4 k}{i(2\pi)^4}
\frac{k^2-4a_0^2+2k_3^2}{(k^2-4a_0^2)^2-16a_0^2 k_3^2}=0 \ . 
\eeq
From Eqs.(\ref{3-21}) and (\ref{3-22}), the inverse propagator for rho meson with $\mu=3$ and $i\neq 3$ 
can be rewritten as 
\beq\label{3-26}
\Gamma_{\rho,i\neq 3}^{33}(p\rightarrow 0)=\frac{2}{G}-4\int\!\!\frac{d^4 k}{i(2\pi)^4}
{\rm tr}\left[
\gamma^{3}\tau_i\frac{1}{\fsl{k}-2a_0\gamma_5\gamma^3\tau_3}\gamma^3\tau_i
\frac{1}{\fsl{k}-2a_0\gamma_5\gamma^3\tau_3}\right]\ . 
\eeq
Similarly, from Eq.(\ref{3-22}), the inverse propagator for $a$ meson with $\mu=3$ and $i\neq 3$ can be rewritten as 
\beq\label{3-27}
\Gamma_{a,i\neq 3}^{33}(p\rightarrow 0)
&=&\frac{2}{G}
-4\int\!\!\frac{d^4 k}{i(2\pi)^4}{\rm tr}\left[
\gamma_5\gamma^{3}\tau_i\frac{1}{\fsl{k}-2a_0\gamma_5\gamma^3\tau_3}
\gamma_5\gamma^3\tau_i
\frac{1}{\fsl{k}-2a_0\gamma_5\gamma^3\tau_3}\right]\nonumber\\
&=&\frac{2}{G}
-4\int\!\!\frac{d^4 k}{i(2\pi)^4}{\rm tr}\left[
\gamma^{3}\tau_i\frac{1}{\fsl{k}-2a_0\gamma_5\gamma^3\tau_3}
\gamma^3\tau_i
\frac{1}{\fsl{k}-2a_0\gamma_5\gamma^3\tau_3}\right]\ , 
\eeq
where we used the anticommutation relation $\{\ \gamma^\mu\ , \ \gamma_5\ \}=0$ and $\gamma_5^2=1$.   
Thus, the same expressions for the inverse propagators of  rho meson with $\mu=3$ and $i\neq 3$ 
and $a$ meson with $\mu=3$ and $i\neq 3$ are obtained. 
After some calculations described in Appendix A, we finally obtain $\Gamma_{\rho,i\neq 3}^{33}(p\rightarrow 0)$ 
and $\Gamma_{a,i\neq 3}^{33}(p\rightarrow 0)$ as 
\beq\label{3-28}
{{\Gamma^{33}}_{\rho/a,i\neq 3}}(p\rightarrow 0)
&=&\frac{2}{G}-96{\int}\frac{d^4k}{i(2\pi)^4}\frac{k^2-4{a_0}^2+2{k_3}^2}{\left[(k^2-4{a_0}^2)^2-16{a_0}^2{k_3}^2\right]}\ .
\eeq
By using the gap equation in Eq.(\ref{3-25}), 
\beq\label{3-29}
{{\Gamma^{33}}_{\rho/a,i\neq 3}}(p\rightarrow 0) = 0 
\eeq
is obtained. 
Because ${{\Gamma^{33}}_{\rho/a,i\neq 3}}(p)$ corresponds to the inverse propagator $p^2-m_{\rho/a}^2$, 
thus $\rho^3_{i=1\ {\rm or}\ 2}$ and $a^{3}_{i=1\ {\rm or}\ 2}$ mesons have no masses. 
Therefore, these mesons are the massless bosons. 
We can check that the other mesons have non-zero masses.

\subsection{The case of $\sigma_s(=2\sigma_0) \neq 0$ and $\sigma_{pv}(=2a_0)\neq 0$}

In the case of $\sigma_0\neq 0$ and $a_0\neq 0$, the gap equations are nothing but Eq.(\ref{2-15}) : 
\beq\label{3-30}
& &
1+48G_S\int \frac{d^4 k}{i(2\pi)^4}
\frac{k^2-4\sigma_0^2+4a_0^2}{(k^2-4\sigma_0^2-4a_0^2)^2-16a_0^2(k_3^2+4\sigma_0^2)}=0\ , \nonumber\\
& &
1-48G\int \frac{d^4 k}{i(2\pi)^4}
\frac{k^2+4\sigma_0^2-4a_0^2+2k_3^2}{(k^2-4\sigma_0^2-4a_0^2)^2-16a_0^2(k_3^2+4\sigma_0^2)}=0\ . 
\eeq
First, let us consider the pion inverse propagator in Eq.(\ref{3-20}). 
Since $\tau_3$ is already included in $\Gamma_{\pi,i}(p)$, let us consider 
$\Gamma_{\pi,i=3}(p)$. 
Then, we obtain 
\beq\label{3-31}
& &\Gamma_{\pi,i=3}(p\rightarrow 0)\nonumber\\
&=&
-\frac{2}{G_S}+4\int\!\!\frac{d^4 k}{i(2\pi)^4}
{\rm tr}\left[
\gamma_5\frac{1}{\fsl{k}-2\sigma_0-2a_0\gamma_5\gamma^3\tau_3} \gamma_5
\frac{1}{\fsl{k}-2\sigma_0-2a_0\gamma_5\gamma^3\tau_3}\right] ,\ \  \ \ 
\eeq
where $\tau_3^2=1$ is used. 
The detailed calculation is described in Appendix A. 
After some calculations, we easily obtain $\Gamma_{\pi,i=3}(p\rightarrow 0)$ as 
\beq\label{3-32}
\Gamma_{\pi,i=3}(p\rightarrow 0)
&=&-\frac{2}{G_S}
-96{\int}{\frac{d^4x}{i(2\pi)^4}}\frac{k^2-4{\sigma_0}^2+4{a_0}^2}{(k^2-4{\sigma_0}^2-4{a_0}^2)^2
-16{a_0}({k_3}^2+4{\sigma_0}^2)}\nonumber\\
&=&0 \ , 
\eeq
where we used the first gap equation in Eq.(\ref{3-30}). 
Thus, the third component of pion, $\pi_3$, is the massless boson. 

Secondly, let us consider the inverse propagator of $a$ meson with $\mu=3$ and $i\neq 3$ in Eq.(\ref{3-22}). 
By using the anticommutation relations for the Dirac and the isospin matrices, we first obtain 
$\Gamma^{33}_{a,i\neq 3}(p\rightarrow 0)$ as 
\beq\label{3-33}
& &\Gamma^{33}_{a,i\neq 3}(p\rightarrow 0) \nonumber\\
&=&\frac{2}{G}-4{\int}\frac{d^4k}{i(2\pi)^4}\mathrm{tr}
\left[{\gamma^3}\frac{1}{\fsl{k}+2{\sigma_0}+2{a_0}{\gamma_5}{\gamma^3}{\tau_3}}{\gamma^3}
\frac{1}{\fsl{k}-2\sigma_0-2{a_0}{\gamma_5}{\gamma^3}{\tau_3}}\right]
\eeq
After tedious calculation described in Appendix A, finally, we obtain the inverse propagator 
$\Gamma^{33}_{a,i\neq 3}(p\rightarrow 0)$ as 
\beq\label{3-34}
\Gamma^{33}_{a,i\neq 3}(p\rightarrow 0)
&=&\frac{2}{G}
-96{\int}\frac{d^4k}{i(2\pi)^4}
\frac{k^2+4{\sigma_0}^2-4{a_0}^2+2{k_3}^2}{(k^2-4{\sigma_0}^2-4{a_0}^2)^2-16{a_0}({k_3}^2+4{\sigma_0}^2)}\nonumber\\
&=&0\ , 
\eeq
where we used the second gap equation for $a_0$ in Eq.(\ref{3-30}). 
Thus, $a$ mesons with $\mu=3$ and $i=1,2$, $a^3_{i=1\ {\rm or}\ 2}$, are the massless bosons adding to $\pi_3$.

Now, we have three NG modes. As will be discussed in section 6, 
in the case with both the chiral condensate and the pseudovector condensate, five generators of the 
chiral $su_V(2)\times su_A(2)$ symmetry should be broken. 
Thus, two NG modes are missing. 
Here, let us show that the pion mode, $\pi_1$ and the rho meson mode $\rho_2^{\mu=3}$ mix. 
Also, it will be shown that the pion mode, $\pi_2$ and the rho meson mode $\rho_1^{\mu=3}$ mix. 

Here, the two-point vertex function in Eq.(\ref{3-17}) is rewritten as
\beq\label{3-35}
\Gamma_{\alpha\beta}(p)(2\pi)^4\delta^4(p+q)
=\left. \frac{\delta^2 \Gamma}{\delta \alpha (p) \delta \beta(q)}\right|_{\sigma=\sigma_0,a_3^3=a_0,\bm{\pi}=\bm{\rho}_{\mu}
=\bm{a}_{\mu\neq 3}=a_{\mu=3}^{i\neq 3}=0}\ ,
\eeq
where $\Delta_\alpha^{-1}(p)=-\Gamma_{\alpha\alpha}(p)$. 
Then, the two-point vertex functions for $\alpha$ or $\beta=\pi_1,\ \pi_2,\ \rho_1^{\mu=3},\ \rho_2^{\mu=3}$ are 
similarly calculated as 
\beq\label{3-36}
\Gamma_{\pi_1\pi_1}(p\rightarrow 0)&=&\Gamma_{\pi_2\pi_2}(p\rightarrow 0)\nonumber\\
&=&-\frac{2}{G_S}-\int\frac{d^4k}{i(2\pi)^4}\mathrm{tr}
\Biggl[(-2i\gamma_5\tau_{1/2})\frac{1}{\fsl{k}-2{\sigma_0}-2{a_0}{\gamma_5}{\gamma^3}{\tau_3}}
\nonumber\\
& &\qquad\qquad\qquad\qquad\quad
\times
(-2i\gamma_5\tau_{1/2})
\frac{1}{\fsl{k}-2\sigma_0-2{a_0}{\gamma_5}{\gamma^3}{\tau_3}}\Biggl]\nonumber\\
&=&-\frac{2}{G_S}-96\int\frac{d^4k}{i(2\pi)^4}\frac{k^2-4\sigma_0^2-4a_0^2}{(k^2-4\sigma_0^2-4a_0^2)^2-16a_0^2(k_3^2+4\sigma_0^2)}\nonumber\\
&=&96\int \frac{d^4k}{i(2\pi)^4}\frac{8a_0^2}{(k^2-4\sigma_0^2-4a_0^2)^2-16a_0^2(k_3^2+4\sigma_0^2)}\ . 
\eeq
Here, the gap equation (\ref{3-30}) is used to obtain the last equality. 
Similarly, we obtain $\Gamma_{\rho_2^{\mu=3}\rho_2^{\mu=3}}$ as 
\beq\label{3-37}
\Gamma_{\rho_2^{\mu=3}\rho_2^{\mu=3}}(p\rightarrow 0)&=&\Gamma_{\rho_1^{\mu=3}\rho_1^{\mu=3}}(p\rightarrow 0)\nonumber\\
&=&\frac{2}{G}-\int\frac{d^4k}{i(2\pi)^4}\mathrm{tr}
\Biggl[(-2\gamma^3\tau_{2/1})\frac{1}{\fsl{k}-2{\sigma_0}-2{a_0}{\gamma_5}{\gamma^3}{\tau_3}}
\nonumber\\
& &\qquad\qquad\qquad\qquad\quad
\times
(-2\gamma^3\tau_{2/1})
\frac{1}{\fsl{k}-2\sigma_0-2{a_0}{\gamma_5}{\gamma^3}{\tau_3}}\Biggl]\nonumber\\
&=&\frac{2}{G}-96\int\frac{d^4k}{i(2\pi)^4}\frac{k^2+2k_3^2-4\sigma_0^2-4a_0^2}{(k^2-4\sigma_0^2-4a_0^2)^2-16a_0^2(k_3^2+4\sigma_0^2)}\nonumber\\
&=&96\int \frac{d^4k}{i(2\pi)^4}\frac{8\sigma_0^2}{(k^2-4\sigma_0^2-4a_0^2)^2-16a_0^2(k_3^2+4\sigma_0^2)}\ ,
\eeq
where the gap equation (\ref{3-30}) is used to obtain the last equality. 
It should be noted that, for $\pi_{1/2}$ and $\rho_{2/1}$ modes, the off-diagonal elements of the two-point vertex functions exist. 
From (\ref{3-35}), the off-diagonal elements are calculated, for example $\alpha=\pi_a$ and $\beta=\rho_b^{\mu=3}$, as 
\beq
\Gamma_{\pi_1\rho_2^{\mu=3}}(p\rightarrow 0)&=&
-4\int\frac{d^4k}{i(2\pi)^4}\mathrm{tr}
\Biggl[(-2i\gamma_5\tau_{1})\frac{1}{\fsl{k}-2{\sigma_0}-2{a_0}{\gamma_5}{\gamma^3}{\tau_3}}
\nonumber\\
& &\qquad\qquad\qquad\qquad\quad
\times
(-2\gamma^3\tau_{2})
\frac{1}{\fsl{k}-2\sigma_0-2{a_0}{\gamma_5}{\gamma^3}{\tau_3}}\Biggl]\nonumber\\
&=&96\int \frac{d^4k}{i(2\pi)^4}\frac{-8\sigma_0a_0}{(k^2-4\sigma_0^2-4a_0^2)^2-16a_0^2(k_3^2+4\sigma_0^2)}\nonumber\\
&=&\Gamma_{\rho_2^{\mu=3}\pi_1}(p\rightarrow 0)\ , 
\label{3-38}\\
\Gamma_{\pi_2\rho_1^{\mu=3}}(p\rightarrow 0)&=&
-4\int\frac{d^4k}{i(2\pi)^4}\mathrm{tr}
\Biggl[(-2i\gamma_5\tau_{2})\frac{1}{\fsl{k}-2{\sigma_0}-2{a_0}{\gamma_5}{\gamma^3}{\tau_3}}
\nonumber\\
& &\qquad\qquad\qquad\qquad\quad
\times
(-2\gamma^3\tau_{1})
\frac{1}{\fsl{k}-2\sigma_0-2{a_0}{\gamma_5}{\gamma^3}{\tau_3}}\Biggl]\nonumber\\
&=&96\int \frac{d^4k}{i(2\pi)^4}\frac{8\sigma_0a_0}{(k^2-4\sigma_0^2-4a_0^2)^2-16a_0^2(k_3^2+4\sigma_0^2)}\nonumber\\
&=&\Gamma_{\rho_1^{\mu=3}\pi_2}(p\rightarrow 0)\ . 
\label{3-39}
\eeq 
The other off-diagonal elements are equal to zero due to the property of the traceless : 
\beq\label{3-40}
& &\Gamma_{\pi_1\pi_2}(p\rightarrow 0)=\Gamma_{\pi_2\pi_1}(p\rightarrow 0)=\Gamma_{\pi_1\rho_1^{\mu=3}}(p\rightarrow 0)
=\Gamma_{\rho_1^{\mu=3}\pi_1}(p\rightarrow 0)\nonumber\\
&=&\Gamma_{\rho_2^{\mu=3}\pi_2}(p\rightarrow 0)=\Gamma_{\pi_2\rho_2^{\mu=3}}(p\rightarrow 0)
=\Gamma_{\rho_2^{\mu=3}\rho_1^{\mu=3}}(p\rightarrow 0)=\Gamma_{\rho_1^{\mu=3}\rho_2^{\mu=3}}(p\rightarrow 0)
\nonumber\\
&=&0 \ .
\eeq
Thus, the mass matrix derived from the two-point vertex functions is written as 
\beq\label{3-41}
& &
\left(
\begin{array}{llll}
\Gamma_{\pi_1\pi_1}(p\rightarrow 0) & \Gamma_{\pi_1\rho_2^{\mu=3}}(p\rightarrow 0) 
& \Gamma_{\pi_1\pi_2}(p\rightarrow 0) & \Gamma_{\pi_1\rho_1^{\mu=3}}(p\rightarrow 0) \\
\Gamma_{\rho_2^{\mu=3}\pi_1}(p\rightarrow 0) & \Gamma_{\rho_2^{\mu=3}\rho_2^{\mu=3}}(p\rightarrow 0) 
& \Gamma_{\rho_2^{\mu=3}\pi_2}(p\rightarrow 0) & \Gamma_{\rho_2^{\mu=3}\rho_1^{\mu=3}}(p\rightarrow 0) \\
\Gamma_{\pi_2\pi_1}(p\rightarrow 0) & \Gamma_{\pi_2\rho_2^{\mu=3}}(p\rightarrow 0) 
& \Gamma_{\pi_2\pi_2}(p\rightarrow 0) & \Gamma_{\pi_2\rho_1^{\mu=3}}(p\rightarrow 0) \\
\Gamma_{\rho_1^{\mu=3}\pi_1}(p\rightarrow 0) & \Gamma_{\rho_1^{\mu=3}\rho_2^{\mu=3}}(p\rightarrow 0) 
& \Gamma_{\rho_1^{\mu=3}\pi_2}(p\rightarrow 0) & \Gamma_{\rho_1^{\mu=3}\rho_1^{\mu=3}}(p\rightarrow 0)
\end{array}
\right)\nonumber\\
&=&K\left(
\begin{array}{cccc}
8a_0^2 & -8\sigma_0a_0 & 0 & 0 \\
-8\sigma_0a_0 & 8\sigma_0^2 & 0 & 0\\
0 & 0 & 8a_0^2 & 8\sigma_0a_0 \\
0 & 0 & 8\sigma_0 a_0 & 8\sigma_0^2
\end{array}
\right) \ , 
\eeq
where
\beq
K=96\int \frac{d^4k}{i(2\pi)^4}\frac{1}{(k^2-4\sigma_0^2-4a_0^2)^2-16a_0^2(k_3^2+4\sigma_0^2)}\ . 
\eeq
It is seen from Eq.(\ref{3-41}) that the mode mixing occurs between $\pi_1$ and $\rho_2$ and/or $\pi_2$ and $\rho_1$, 
respectively, because the mass matrix becames to the block diagonal.

As fo the mixing between $\pi_1$ and $\rho_2$, namely 
\beq\label{3-42}
& &
M^2_{\pi_1\rho_2^{\mu=3}}=\left(
\begin{array}{ll}
\Gamma_{\pi_1\pi_1}(p\rightarrow 0) & \Gamma_{\pi_1\rho_2^{\mu=3}}(p\rightarrow 0) 
\\
\Gamma_{\rho_2^{\mu=3}\pi_1}(p\rightarrow 0) & \Gamma_{\rho_2^{\mu=3}\rho_2^{\mu=3}}(p\rightarrow 0) 
\end{array}
\right)
=K\left(
\begin{array}{cc}
8a_0^2 & -8\sigma_0a_0 \\
-8\sigma_0a_0 & 8\sigma_0^2 
\end{array}
\right) \ , 
\eeq
we can diagonalize the above mass matrix by using the unitary matrix $U$ as 
\beq\label{3-43}
& &U=\frac{1}{\sqrt{\sigma_0^2+a_0^2}}\left(
\begin{array}{cc}
\sigma_0 & a_0 \\
a_0 & -\sigma_0 
\end{array}
\right) \ , \qquad
U^{-1}=U^{\dagger}=\frac{1}{\sqrt{\sigma_0^2+a_0^2}}\left(
\begin{array}{cc}
\sigma_0 & a_0 \\
a_0 & -\sigma_0 
\end{array}
\right) \ , \nonumber\\
& &
UM^2_{\pi_1\rho_2^{\mu=3}}U^{-1}=
\left(
\begin{array}{cc}
0 & \quad 0 \\
0 &\quad 8K(\sigma_0^2+a_0^2) 
\end{array}
\right) \ . 
\eeq
Thus, we conclude that the 
$(\sigma_0\pi_1+a_0\rho_2^{\mu=3})/\sqrt{\sigma_0^2+a_0^2}$ is the massless NG mode and 
$(a_0\pi_1-\sigma_0\rho_2^{\mu=3})/\sqrt{\sigma_0^2+a_0^2}$ is the massive mode with 
the mass $\sqrt{8K(\sigma_0^2+a_0^2)}$.

Secondly, as fo the mixing between $\pi_2$ and $\rho_1$, namely 
\beq\label{3-44}
& &
M^2_{\pi_2\rho_1^{\mu=3}}=\left(
\begin{array}{ll}
\Gamma_{\pi_2\pi_2}(p\rightarrow 0) & \Gamma_{\pi_2\rho_1^{\mu=3}}(p\rightarrow 0) 
\\
\Gamma_{\rho_1^{\mu=3}\pi_2}(p\rightarrow 0) & \Gamma_{\rho_1^{\mu=3}\rho_1^{\mu=3}}(p\rightarrow 0) 
\end{array}
\right)
=K\left(
\begin{array}{cc}
8a_0^2 & 8\sigma_0a_0 \\
8\sigma_0a_0 & 8\sigma_0^2 
\end{array}
\right) \ ,  
\eeq
the mass matrix can be diagonalized by using the unitary matrix $U$ as 
\beq\label{3-45}
& &U=\frac{1}{\sqrt{\sigma_0^2+a_0^2}}\left(
\begin{array}{cc}
a_0 & \sigma_0 \\
\sigma_0 & -a_0 
\end{array}
\right) \ , \qquad
U^{-1}=U^{\dagger}=\frac{1}{\sqrt{\sigma_0^2+a_0^2}}\left(
\begin{array}{cc}
a_0 & \sigma_0 \\
\sigma_0 & -a_0 
\end{array}
\right) \ , \nonumber\\
& &
UM^2_{\pi_2\rho_1^{\mu=3}}U^{-1}=
\left(
\begin{array}{cc}
8K(\sigma_0^2+a_0^2)  \quad & 0 \\
0 \quad & 0 
\end{array}
\right) \ . 
\eeq
Thus, we conclude that the 
$(\sigma_0\pi_2-a_0\rho_1^{\mu=3})/\sqrt{\sigma_0^2+a_0^2}$ is the massless NG mode and 
$(a_0\pi_2+\sigma_0\rho_1^{\mu=3})/\sqrt{\sigma_0^2+a_0^2}$ is the massive mode with 
the mass $\sqrt{8K(\sigma_0^2+a_0^2)}$. 

Thus, we have five NG modes in the case of the chiral symmetry breaking due to both the chiral condensate 
$\sigma_0\ (=\langle \sigma \rangle=-G_S\langle {\bar \psi}\psi\rangle)$ and 
the pseudovector condensate $a_0\ (=\langle a_3^{\mu=3}\rangle=-G\langle \psi^{\dagger}\Sigma_3\psi\rangle)$, namely 
\beq\label{3-46}
a_1^{\mu=3}\ ,\quad a_2^{\mu=3}\ , \quad \pi_3\ , \quad \frac{\sigma_0\pi_1+a_0\rho_2^{\mu=3}}{\sqrt{\sigma_0^2+a_0^2}}\ , \quad
\frac{\sigma_0\pi_2-a_0\rho_1^{\mu=3}}{\sqrt{\sigma_0^2+a_0^2}} \ . \nonumber\\
& &
\eeq

\section{Gap equation with respect to the tensor condensate and the massless modes on the tensor condensate}

\subsection{Gap equation for the tensor condensate}

Let us introduce the NJL model with tensor interaction, instead of the pseudovector interaction, 
which retains the chiral 
symmetry\cite{NJL,Klevansky,HK,Buballa}: 
\begin{equation}\label{4-1}
 \mathcal{L}_{\rm NJL, T}=\bar{\psi}i\gamma^\mu\partial_\mu\psi+G_S[(\bar{\psi}\psi)^2+(\bar{\psi}i\gamma_5\bm{\tau}\psi)^2]
  +G_T[(\bar{\psi} \gamma^\mu\gamma^\nu\bm{\tau}\psi)^2+(\bar{\psi}i\gamma_5\gamma^\mu\gamma^\nu\psi)^2] \ .
\end{equation}
In a similar way developed in \S 2, we use the auxiliary field method and introduce the functional Gaussian integral leading to unit as 
\beq\label{4-2}
& &1=\int{\cal D}\sigma'{\cal D}{\bm{\pi}}'\ \exp\left(i\int d^4x \left[-\frac{1}{G_S}
\left(\sigma'{}^2+\bm{\pi}'{}^2\right)\right]\right)\ , \nonumber\\
& &1=\int{\cal D}\bm{t}'{}^{\mu\nu}{\cal D}{p}'{}^{\mu\nu}\ \exp\left(i\int d^4x \left[-\frac{1}{G_T}
\left(\bm{t}'{}^{\mu\nu}\cdot{\bm{t}}'_{\mu\nu}+{p}'{}^{\mu\nu} {p}'_{\mu\nu}\right)\right]\right)\ . 
\eeq
Inserting the above unit into the generating functional $Z$ in Eq.(\ref{2-2}) and 
replacing $\sigma'$, $\bm{\pi}'$, $\bm{t}'{}^{\mu\nu}$ and ${p}'{}^{\mu\nu}$ into 
$\sigma$, $\bm{\pi}$, $\bm{t}{}^{\mu\nu}$ and ${p}{}^{\mu\nu}$ as 
\beq\label{4-3}
& &\sigma'=\sigma +G_S(\bar{\psi}\psi) \ , \nonumber\\
& &\bm{\pi}' = \bm{\pi}+G_S(\bar{\psi}i\gamma_5\bm{\tau}\psi) \ , \nonumber\\
& &\bm{t}'{}^{\mu\nu}=\bm{t}^{\mu\nu}+G_T(\bar{\psi}\gamma^\mu\gamma^\nu\bm{\tau}\psi) \ , \nonumber\\ 
& &{p}'{}^{\mu\nu} ={p}^{\mu\nu}+G_T(\bar{\psi}i\gamma_5\gamma^\mu\gamma^\nu\psi) \ ,
\eeq
the generating functional is rewritten as 
\beq\label{4-4}
Z&=&\int{\cal D}\psi{\cal D}{\bar{\psi}}\int{\cal D}\sigma{\cal D}{\bm{\pi}}{\cal D}\bm{t}^{\mu\nu}{\cal D}{{p}}^{\mu\nu}
\exp \left(i\int d^4x {\wtilde{\cal L}}\right)\ ,
\eeq
where 
\beq\label{4-5}
{\wtilde {\cal L}}&=&
\bar{\psi}(i\gamma^\mu\partial_\mu - 2\sigma - 2\bm{\pi}\cdot i\gamma_5\bm{\tau} 
                     - 2\bm{t}_{\mu\nu}\cdot\gamma^\mu\gamma^\nu\bm{\tau} - 2{p}_{\mu\nu} i\gamma_5\gamma^{\mu\nu}
)\psi 
\nonumber\\
& & - \frac{1}{G_S}(\sigma^2+\bm{\pi}^2) 
                     - \frac{1}{G_T}(\bm{t}^{\mu\nu}\cdot\bm{t}_{\mu\nu}+{p}^{\mu\nu}{p}_{\mu\nu})\ .
\eeq
Here, from $\delta {\wtilde {\cal L}}/\delta \sigma=0$ and so on, we obtain 
\beq\label{4-6}
& &\sigma =-G_S(\bar{\psi}\psi) \ , \nonumber\\
& &\bm{\pi}=-G_S(\bar{\psi}i\gamma_5\bm{\tau}\psi) \ , \nonumber\\
& &\bm{t}_{\mu\nu}=-G_T(\bar{\psi}\gamma_{\mu}\gamma_{\nu}\bm{\tau}\psi) \ , \nonumber\\ 
& &{p}_{\mu\nu}=-G_T(\bar{\psi}i\gamma_5\gamma_\mu\gamma_\nu\psi) \ .
\eeq
Further, integrating $\psi$ and ${\bar \psi}$, the generating functional can be expressed as 
\beq\label{4-7}
Z&=&\int{\cal D}\sigma{\cal D}{\bm{\pi}}{\cal D}\bm{t}^{\mu\nu}{\cal D}{{p}}^{\mu\nu}
\exp \left(i\Gamma\right)\ ,\nonumber\\
\Gamma&=&\int d^4 x\left[-\frac{1}{G_S}(\sigma^2+\bm{\pi}^2) 
                     - \frac{1}{G_T}(\bm{t}^{\mu\nu}\cdot\bm{t}_{\mu\nu}+{p}^{\mu\nu}{p}_{\mu\nu})\right]
\nonumber\\
& &\qquad\quad 
-i\ln {\rm Det} \left(i\gamma^\mu\partial_\mu - 2\sigma - 2\bm{\pi}\cdot i\gamma_5\bm{\tau} 
                     - 2\bm{t}_{\mu\nu}\cdot\gamma^\mu\gamma^\nu\bm{\tau} - 2{p}_{\mu\nu} i\gamma_5\gamma^\mu\gamma^\nu\right)
\ . \nonumber\\
& &
\eeq
It should be noted that the tensor mode $\bm{t}^{12}$ is written as 
\beq\label{4-8} 
\bm{t}^{12}&=&-G_T\ \bar{\psi}\gamma^1\gamma^2\bm{\tau}\psi \nonumber\\
&=&iG_T\ \bar{\psi}\Sigma_3 \bm{\tau} \psi\ .
\eeq
Here, we use the Dirac representation for the Dirac gamma matrices, namely, 
\beq\label{4-9}
& &\gamma^1\gamma^2 = -i\Sigma_3=-i\begin{pmatrix}
		\sigma_3 & 0  \\
		 0 & \sigma_3 
	\end{pmatrix} \ . 
\eeq

Hereafter, let us assume that there exist the chiral condensate $\sigma_0=\braket{\sigma}$ 
and tensor condensate $t_0=\braket{t^{\mu=1\ \nu=2}_{i=3}}(=-\braket{t^{\mu=2\ \nu=1}_{i=3}})$, 
namely 
\beq\label{4-10}
& &\sigma_0\equiv \braket{\sigma}=-G_S\braket{{\bar \psi}\psi}\ , \nonumber\\
& &t_0\equiv \braket{t_{\mu=1\mu=2}^{i=3}}=iG_T\braket{{\bar \psi}\Sigma_3\tau_3 \psi}\ . 
\eeq
Under these quark-antiquark condensates, the effective potential $V(\sigma_0,t_0)$ can be 
derived from the effective action $\Gamma$ as
\beq\label{4-11}
\Gamma(\sigma=\sigma_0,t_3^{12}=t_0)&=&-V(\sigma_0,t_0)\int d^4 x\ , \nonumber\\
V(\sigma_0,t_0)
&=&\frac{1}{G_S}\sigma_0^2+\frac{2}{G_T}t_0^2-
\int\frac{d^4 k}{i(2\pi)^4}\ln\det \left[
\gamma^\mu k_\mu -2\sigma_0-4t_0\gamma^1\gamma^2\tau_3\right]\nonumber\\
&=&\frac{1}{G_S}\sigma_0^2+\frac{2}{G_T}t_0^2-
\int\frac{d^4 k}{i(2\pi)^4}{\rm tr}\ln \left[
\gamma^\mu k_\mu -2\sigma_0-4t_0\gamma^1\gamma^2\tau_3\right]\ , \nonumber\\
& &  
\eeq
where trace is taken by the Dirac gamma matrices, isospin space and color space. 
Here, it should be noted that $t^{\mu=2\ \nu=1}_{i=3}=-t^{\mu=1\ \nu=2}_{i=3}=-t_0$ is used.

Hereafter, we consider the tensor condensate $t_0$ only because both the tensor condensate and the chiral condensate 
$\sigma_0$ may not coexist according to our previous work.\cite{Matsuoka2016} 
Thus, we set $\sigma_0=0$ in Eq. (\ref{4-11}). 
The condensate $t_0$ is determined by the following gap equation : 
\beq\label{4-13}
& &\frac{\partial V}{\partial t_0}
=\frac{4}{G_T}t_0
-\int \frac{d^4 k}{i(2\pi)^4}
{\rm tr} \left[\frac{1}{\gamma^\mu k_\mu -4t_0\gamma^1\gamma^2\tau_3}(-4\gamma^1\gamma^2\tau_3)\right]=0\ , 
\eeq 
which, after some tedious calculation for the gamma matrices, leads to 
\beq\label{4-14}
& &t_0\left[
1-32G_T\int \frac{d^4 k}{i(2\pi)^4}
\frac{k^2+16t_0^2+2(k_1^2+k_2^2)}{(k^2+16t_0^2)^2+64t_0^2(k_1^2+k_2^2)}\right]=0\ , 
\eeq
where $k^2=k_0^2-\bm{k}^2$, and $k_1$ and $k_2$ mean the first and the second components of the momentum vector $\bm{k}$.

\subsection{Massless modes on the tensor condensate $t_0$}

Thus, two-point vertex functions for $\alpha=\bm{t}^{\mu\nu}$ and $p^{\mu\nu}$ in Eq.(\ref{3-17}) are obtained 
by using $\Gamma$ in Eq.(\ref{4-7}) as 
\beq
\Gamma_{t,i}^{\mu\nu,\rho\sigma}(p)
&=&-\frac{4}{G_T}(g^{\mu\rho}g^{\nu\sigma}-g^{\mu\sigma}g^{\nu\rho})\nonumber\\
& &-\int\!\!\frac{d^4 k}{i(2\pi)^4}
{\rm tr}\left[
(-4\gamma^{\mu}\gamma^{\nu}\tau_i)\frac{1}{\fsl{k}-4t_0\gamma^1\gamma^2\tau_3}\right. 
\left. \times (-4\gamma^\rho\gamma^{\sigma}\tau_i)
\frac{1}{\fsl{k}+\fsl{p}-4t_0\gamma^1\gamma^2\tau_3}\right]\ , \nonumber\\
& &
\label{4-15}\\
\Gamma_{p,i}^{\mu\nu,\rho\sigma}(p)
&=&-\frac{4}{G_T}(g^{\mu\rho}g^{\nu\sigma}-g^{\mu\sigma}g^{\sigma\rho})\nonumber\\
& &-\int\!\!\frac{d^4 k}{i(2\pi)^4}
{\rm tr}\left[
(-4i\gamma_5\gamma^{\mu}\gamma^{\nu})\frac{1}{\fsl{k}-4t_0\gamma^1\gamma^2\tau_3} \right. 
\left. \times (-4i\gamma_5\gamma^\rho\gamma^\sigma)
\frac{1}{\fsl{k}+\fsl{p}-4t_0\gamma^1\gamma^2\tau_3}\right]\ , \nonumber\\
& &
\label{4-16}
\eeq
where we used the relations $\partial(t^{\alpha\beta}t_{\alpha\beta})/\partial t_{\mu\nu}
=\partial(t^{\mu\nu}t_{\mu\nu}+t^{\nu\mu}t_{\nu\mu})/\partial t_{\mu\nu}=4t^{\mu\nu}$ and 
$\partial t^{\mu\nu}t_{\alpha\beta}/\partial t_{\rho\sigma}=g^{\mu\rho}g^{\nu\sigma}-g^{\mu\sigma}g^{\nu\rho}$.

Here, we only consider the case with $\sigma_0=0$ and $t_0\neq 0$. 
In this case, the gap equation in Eq.(\ref{4-14}) can be expressed as 
\beq\label{4-17}
1-32G_T\int \frac{d^4 k}{i(2\pi)^4}
\frac{k^2+16t_0^2+2(k_1^2+k_2^2)}{(k^2+16t_0^2)^2+64t_0^2(k_1^2+k_2^2)}=0 \ . 
\eeq
From Eq.(\ref{4-15}), the inverse propagator with $\mu=\rho=1$, $\nu=\sigma=2$ and $i\neq 3$ 
can be rewritten as 
\beq\label{4-18}
\Gamma_{t,i\neq 3}^{12,12}(p\rightarrow 0)
&=&-\frac{4}{G_T}-16\int\!\!\frac{d^4 k}{i(2\pi)^4}
{\rm tr}\left[
\gamma^{1}\gamma^2\tau_i\frac{1}{\fsl{k}-4t_0\gamma^1\gamma^2\tau_3}\gamma^1\gamma^2\tau_i
\frac{1}{\fsl{k}-4t_0\gamma^1\gamma^2\tau_3}\right]\nonumber\\
&=&-\frac{4}{G_T}-16\int\!\!\frac{d^4 k}{i(2\pi)^4}
{\rm tr}\left[
\gamma^{1}\gamma^2\frac{1}{\fsl{k}+4t_0\gamma^1\gamma^2\tau_3}\gamma^1\gamma^2
\frac{1}{\fsl{k}-4t_0\gamma^1\gamma^2\tau_3}\right]\ , \nonumber\\
& & 
\eeq
where we used the anticommutation relation $\{\ \tau_i\ , \ \tau_3\ \}=0$ with $i=1,2$ and $\tau_i^2=1$. 
Similarly, from Eq.(\ref{4-16}), the inverse propagator with $\mu=\rho=1$ and $\nu=\sigma=2$ can be rewritten as 
\beq\label{4-19}
\Gamma_{p}^{12,12}(p\rightarrow 0)
&=&-\frac{4}{G_T}
+16\int\!\!\frac{d^4 k}{i(2\pi)^4}{\rm tr}\left[
\gamma^5\gamma^{1}\gamma^2\frac{1}{\fsl{k}-4t_0\gamma^1\gamma^2\tau_3}
\gamma^5\gamma^1\gamma^2
\frac{1}{\fsl{k}-4t_0\gamma^1\gamma^2\tau_3}\right] \nonumber\\
&=&-\frac{4}{G_T}
-16\int\!\!\frac{d^4 k}{i(2\pi)^4}{\rm tr}\left[
\gamma^{1}\gamma^2\frac{1}{\fsl{k}+4t_0\gamma^1\gamma^2\tau_3}
\gamma^1\gamma^2
\frac{1}{\fsl{k}-4t_0\gamma^1\gamma^2\tau_3}\right] \ , \nonumber\\
& & 
\eeq
where we used the anticommutation relation $\{\ \gamma^\mu\ , \ \gamma_5\ \}=0$ and $\gamma_5^2=1$.   
Thus, the same expression for the inverse propagator with $\mu=\rho=1$, $\nu=\sigma=2$ and $i\neq 3$ for 
$t$-mode is obtained.

After tedious calculations similar to the case of the pseudovector condensate $a_0\neq 0$ described in Appendix A, 
we finally obtain $\Gamma_{t,i\neq 3}^{12,12}(p\rightarrow 0)$ 
and $\Gamma_{p}^{12,12}(p\rightarrow 0)$ as 
\beq\label{4-20}
{\Gamma^{12,12}}_{t,i\neq 3/p}(p\rightarrow 0)
&=&-\frac{4}{G_T}+128{\int}\frac{d^4k}{i(2\pi)^4}\frac{k^2+16{t_0}^2+2({k_1}^2+{k_2}^2)}
{\left[(k^2+16{t_0}^2)^2+64{t_0}^2({k_1}^2+{k_2}^2)\right]}\ . \nonumber\\
& &
\eeq
By using the gap equation in Eq.(\ref{4-17}), we obtain 
\beq\label{4-21}
{{\Gamma^{12,12}}_{t,i\neq 3/p}}(p\rightarrow 0) = 0 \ .
\eeq
Because ${{\Gamma^{12,12}}_{t,i\neq 3/p}}(p)$ corresponds to the inverse propagator $p^2-m_{t/p}^2$, 
thus the mesons corresponding to the modes $t^{12}_{i=1\ {\rm or}\ 2}$ and $p^{12}$ have no masses. 
Therefore, these mesons correspond to the massless Nambu-Goldstone bosons. 
Here, from Eq.(\ref{4-6}), $p^{12}\propto {\bar \psi}i\gamma_5\gamma^1\gamma^2\psi 
={\bar\psi}\gamma^0\gamma^3\psi$. 
Thus, $p$-mode corresponds to the isoscalar-tensor mode with $\mu=0$ and $\nu=3$. 
Thus, two isovector-tensor modes with $i=1,2$ and $\mu=1$, $\nu=2$ and 
one isoscalar-tensor mode with $\mu=0$ and $\nu=3$ are the massless Nambu-Goldstone modes. 
We can check that the other meson modes have non-zero masses.

\section{Summary and concluding remarks}

In this paper, the possibility of the existence of the pseudovector condensed phase has been investigate 
in the quark matter with the finite quark chemical potential $\mu$. 
Here, the interaction strength for the pseudovector interaction in the NJL model $G_{pv}(=-G)$ has been 
taken as a free parameter. 
If $G_{pv}\approx G_S$, there is a possibility of the pseudovector condensed phase due to the vacuum effects 
in  a certain region of the quark chemical potential, in which the chiral condensate disappears. 
It has been thus shown that the chiral condensate and the pseudovector condensate do not coexist.
If the pseudovector interaction strength $G_{pv}$ is taken rather large such as $G_{pv}>1.3 G_S$, 
then, the point with $\sigma_{pv}\neq 0$ and $\sigma_s=0$ becomes the absolute minimum of the effective potential, 
while $\sigma_{pv}=0$ and $\sigma_s\neq 0$ is rest as the local minimum. 
However, this situation is unphysical because the vacuum with $\mu=0$ does not reveal 
the quark spin polarization due to the pseudovector condensate
in the realistic case.

Also, the massless meson modes on the pseudovector condensate, which leads to 
the quark spin polarization $\braket{\psi^{\dagger}\Sigma_3\tau_3\psi}(=\braket{{\bar \psi}\gamma_5\gamma_3\tau_3\psi})$, 
have been investigated
in the Nambu-Jona-Lasinio model with the pseudovector interaction between quarks. 
If the spin polarization $\braket{\psi^{\dagger}\Sigma_3\tau_3\psi}$ 
due to the pseudovector condensate $\braket{{\bar \psi}\gamma_5\gamma_3\tau_3\psi}$ 
only exists and the chiral condensate $\braket{{\bar \psi}\psi}$ does not exist, 
four massless modes have appeared, namely, two rho meson modes, $\rho^{\mu=3}_{i=1}$, $\rho^{\mu=3}_{i=2}$ and 
two $a$ meson modes,  $a^{\mu=3}_{i=1}$, $a^{\mu=3}_{i=2}$. 

The reason is as follows : 
Let us consider the chiral $su_A(2)\times su_V(2)$ transformation, namely
\beq\label{6-1}
\psi &\longrightarrow& e^{i\theta_a\gamma_5\tau^a/2}\psi\approx \psi +i\theta_a\gamma_5\frac{\tau^a}{2}\psi\nonumber\\
&\approx& \psi +\theta_a \delta_A^a \psi\ , \qquad {\rm for}\ \ su_A(2)\mathchar`-{\rm transformation} \nonumber\\
\psi &\longrightarrow& e^{i\varphi_a\tau^a/2}\psi\approx \psi +i\varphi_a\frac{\tau^a}{2}\psi\nonumber\\
&\approx& \psi + \varphi_a 
\delta_V^a \psi\ , \qquad {\rm for}\ \ su_V(2)\mathchar`-{\rm transformation} 
\eeq
where $\delta_A^a\psi=i\gamma_5(\tau^a/2)\psi$ and $\delta_V^a\psi=i(\tau^a/2)\psi$, respectively, and $a$ represents 
the isospin indices.  
The Noether currents $j^\mu_a$ and the Noether charge $Q^a$ are calculated as 
\beq
& &j_{A, a}^\mu=\frac{\partial {\cal L}}{\partial (\partial_\mu \psi)}\delta_A^a\psi
=-{\bar \psi}\gamma^\mu\gamma_5\frac{\tau^a}{2}\psi\ , \nonumber\\
& &j_{V, a}^\mu=\frac{\partial {\cal L}}{\partial (\partial_\mu \psi)}\delta_V^a\psi
=-{\bar \psi}\gamma^\mu\frac{\tau^a}{2}\psi\ ,
\label{6-2}\\
& &
Q_A^a=\int d^3{\mib x}\ j_{A,a}^{\mu=0}=-\int d^3{\mib x}\ {\bar \psi}\gamma^0\gamma_5\frac{\tau^a}{2}\psi\ , \nonumber\\
& &
Q_V^a=\int d^3{\mib x}\ j_{V,a}^{\mu=0}=-\int d^3{\mib x}\ {\bar \psi}\gamma^0\frac{\tau^a}{2}\psi\ , 
\label{6-3}
\eeq
Then, under the chiral transformations, the following relations are obtained : 
\beq\label{6-4}
& &[\ iQ_A^a\ , \ {\bar \psi}i\gamma_5\tau^b\psi\ ]
=-\delta^{ab} {\bar \psi}\psi\ , 
\nonumber\\
& &[\ iQ_A^a\ , \ {\bar \psi}\psi\ ]
={\bar \psi} i\gamma_5\tau^a \psi\ , 
\nonumber\\
& &[\ iQ_V^a\ , \ {\bar \psi}i\gamma_5\tau^b\psi\ ]
=2\epsilon^{abc} {\bar \psi}i\gamma_5\tau^c\psi\ , 
\nonumber\\
& &[\ iQ_V^a\ , \ {\bar \psi}\psi\ ]
=0\ . 
\eeq 
Therefore, if the quark-antiquark chiral condensate $\langle {\bar \psi}\psi\rangle$ exists, namely 
 $\langle {\bar \psi}\psi\rangle\neq 0$, then from the first equation in Eq.({\ref{6-4}) ,
\beq\label{6-5}
\langle [\ iQ_A^a\ , \ {\bar \psi}i\gamma_5\tau^b\psi\ ]\rangle 
=-\delta^{ab} \langle {\bar \psi}\psi\rangle \neq 0 
\eeq
is obtained. Here, $\delta^{ab}$ represents the Kronecker delta. 
Thus, the chiral $su_A(2)$-symmetry, whose generators are $Q_A^a\ (a=1,2,3)$, is spontaneously broken and the NG modes are 
\beq\label{6-6}
{\bar \psi}i\gamma_5\tau^b\psi \approx \pi^b\ , 
\eeq
namely, three pions for $b=1,\ 2,\ 3$ are the NG bosons. 
This statement is found in the many textbooks. 

Similarly, keeping in mind about the pseudovector condensate, we obtain 
\beq\label{6-7}
& &[\ iQ_A^a\ , \ {\bar \psi}\gamma^\mu\tau^b\psi\ ]
=-\epsilon^{abc} {\bar \psi}\gamma_5\gamma^\mu\tau^c\psi\ , 
\nonumber\\
& &[\ iQ_A^a\ , \ {\bar \psi}\gamma_5\gamma^\mu\tau^b\psi\ ]
=-\epsilon^{abc}{\bar \psi} \gamma^\mu\tau^c \psi\ , 
\nonumber\\
& &[\ iQ_V^a\ , \ {\bar \psi}\gamma^\mu\tau^b\psi\ ]
=\epsilon^{abc} {\bar \psi}\gamma^\mu\tau^c\psi\ , 
\nonumber\\
& &[\ iQ_V^a\ , \ {\bar \psi}\gamma_5\gamma^\mu\tau^b\psi\ ]
=\epsilon^{abc} {\bar \psi}\gamma_5\gamma^\mu\tau^c\psi \ ,
\eeq 
where $\epsilon^{abc}$ is the completely antisymmetric tensor. 
Therefore, if the pseudovector-type quark-antiquark condensate exists, namely 
$\langle {\bar \psi}\gamma_5\gamma^\mu\tau^3\psi\rangle\neq 0$, then from the first and the fourth equations 
in Eq.({\ref{6-7}) ,
\beq\label{6-8}
& &\langle [\ iQ_A^{a=1}\ , \ {\bar \psi}\gamma^\mu\tau^{b=2}\psi\ ]\rangle
=- \langle {\bar \psi}\gamma_5\gamma^\mu\tau^3\psi\rangle 
\neq 0\ , \nonumber\\
& &\langle [\ iQ_A^{a=2}\ , \ {\bar \psi}\gamma^\mu\tau^{b=1}\psi\ ]\rangle
= \langle {\bar \psi}\gamma_5\gamma^\mu\tau^3\psi\rangle 
\neq 0\ , \nonumber\\
& &\langle[\ iQ_V^{a=1}\ , \ {\bar \psi}\gamma_5\gamma^\mu\tau^{b=2}\psi\ ]\rangle
=\langle {\bar \psi}\gamma_5\gamma^\mu\tau^3\psi\rangle 
\neq 0\ , \nonumber\\
& &\langle[\ iQ_V^{a=2}\ , \ {\bar \psi}\gamma_5\gamma^\mu\tau^{b=1}\psi\ ]\rangle
=-\langle {\bar \psi}\gamma_5\gamma^\mu\tau^3\psi\rangle 
\neq 0
\eeq
are obtained.
Thus, from the first and the second equations, the chiral $su_A(2)$-symmetry, namely $Q_A^1$ and $Q_A^2$,  
is spontaneously broken, which 
leads to the NG modes as
\beq\label{6-9}
& &{\bar \psi}\gamma^\mu \tau^{a=2}\psi \approx \rho^\mu_2\ , \nonumber\\
& &{\bar \psi}\gamma^\mu \tau^{a=1}\psi \approx \rho^\mu_1\ . 
\eeq
Namely, two rho meson modes with $a=1$ and 2 are the NG bosons. 
In addition to the chiral $su_A(2)$-symmetry, the chiral $su_V(2)$-symmetry, namely $Q_V^1$ and $Q_V^2$, is also broken. 
Thus, from the third and the fourth equations in Eq.(\ref{6-8}), we obtain the other two NG modes as 
\beq\label{6-10}
& &{\bar \psi}\gamma_5\gamma^\mu \tau^{a=2}\psi \approx a^\mu_2\ , \nonumber\\
& &{\bar \psi}\gamma_5\gamma^\mu \tau^{a=1}\psi \approx a^\mu_1\ . 
\eeq
Namely, two $a$ meson modes with $a=1$ and 2 are also the NG bosons. 
Therefore, under the pseudovector condensate, the chiral $su_V(2)$-symmetry is broken 
as well as the chiral $su_A(2)$-symmetry breaking.
Thus, four NG modes are realized on the pseudovector condensate. 
This situations are exactly shown in section 4.2.

Next, if the chiral condensate and the pseudovector condensate coexist, namely $\sigma_0\neq 0$ and $a_0\neq 0$, 
the symmetry with respect to $Q_A^1$, $Q_A^2$, $Q_A^3$ is broken due to $\sigma_0\neq 0$ and 
the symmetry for $Q_A^1$, $Q_A^2$, $Q_V^1$ and $Q_V^2$ is also broken due to $a_0\neq 0$. 
Since the symmetry with respect to five generators $Q_A^1$, $Q_A^2$, $Q_A^3$, $Q_V^1$ and $Q_V^2$ is broken, 
it is expected that the five NG modes should appear. 
We already found three massless modes by calculating the diagonal elements of 
the two-point vertex functions $\Gamma_{\alpha}(p)$, namely 
$\Gamma_{\pi, i=3}(p\rightarrow 0)=0$, $\Gamma^{\mu=3,\nu=3}_{a,i\neq 3}(p\rightarrow 0)=0$.  
Here, $\pi^3$ mode related to $Q_A^3$ and $a^{\mu=3}_{i=1\ {\rm and}\ 2}$ modes related to $Q_V^1$ and $Q_V^2$ 
correspond to the NG modes. 
However, the rest two NG modes are naively missing, which related to $Q_A^1$ and $Q_A^2$. 
The generators $Q_A^1$ and $Q_A^2$ appear two times in the symmetry breaking due to $\sigma_0\neq 0$, 
which leads to $\pi_1$ and $\pi_2$ as NG modes, and $a_0\neq 0$, which leads to $\rho_2^{\mu}$ and $\rho_1^{\mu}$ 
as NG modes. 
Thus, $\pi_1$-$\rho_2^{\mu}$ mixing and $\pi_2$-$\rho_1^{\mu}$ mixing may occur. 
Actually, as was shown in section 4, these occurs and 
$(\sigma_0\pi_1+a_0\rho_2^{\mu=3})/\sqrt{\sigma_0^2+a_0^2}$ and 
$(\sigma_0\pi_2-a_0\rho_1^{\mu=3})/{\sqrt{\sigma_0^2+a_0^2}}$ are realized as the massless NG modes, 
while it is well known that, in the case of $\sigma_0\neq 0$, the pion and the $a$ meson mix. \cite{Ulf}

Next, keeping in mind about the tensor condensate only, we obtain 
\beq\label{6-11}
& &[\ iQ_A^a\ , \ {\bar \psi}\gamma^\mu\gamma^\nu\tau^b\psi\ ]
=\delta^{ab} i{\bar \psi}\gamma_5\gamma^\mu\gamma^\nu\psi\ , 
\nonumber\\
& &[\ iQ_A^a\ , \ {\bar \psi}i\gamma_5\gamma^\mu\gamma^\nu\psi\ ]
=-{\bar \psi} \gamma^\mu\gamma^\nu\tau^a \psi\ , 
\nonumber\\
& &[\ iQ_V^a\ , \ {\bar \psi}\gamma^\mu\gamma^\nu\tau^b\psi\ ]
=\epsilon^{abc} {\bar \psi}\gamma^\mu\gamma^\nu\tau^c\psi\ , 
\nonumber\\
& &[\ iQ_V^a\ , \ {\bar \psi}i\gamma_5\gamma^\mu\gamma^\nu\psi\ ]
=0 \ .
\eeq 
Therefore, if the tensor-type quark-antiquark condensate exists, namely 
$\langle {\bar \psi}\gamma^\mu\gamma^\nu\tau^3\psi\rangle\neq 0$, then from the second and the third equations 
in Eq.({\ref{6-11}) ,
\beq\label{6-12}
& &\langle [\ iQ_A^{a=3}\ , \ {\bar \psi}i\gamma_5\gamma^\mu\gamma^\nu\psi\ ]\rangle
=- \langle {\bar \psi}\gamma^\mu\gamma^\nu\tau^3\psi\rangle 
\neq 0\ , \nonumber\\
& &\langle[\ iQ_V^{a=1}\ , \ {\bar \psi}\gamma^\mu\gamma^\nu\tau^{b=2}\psi\ ]\rangle
=\langle {\bar \psi}\gamma^\mu\gamma^\nu\tau^3\psi\rangle 
\neq 0\ , \nonumber\\
& &\langle[\ iQ_V^{a=2}\ , \ {\bar \psi}\gamma^\mu\gamma^\nu\tau^{b=1}\psi\ ]\rangle
=-\langle {\bar \psi}\gamma^\mu\gamma^\nu\tau^3\psi\rangle 
\neq 0
\eeq
are obtained.
Thus, from the first equation, the chiral $su_A(2)$-symmetry is spontaneously broken, namely $Q_A^3$, which 
leads to the NG modes as
\beq\label{6-13}
& &{\bar \psi}i\gamma_5\gamma^\mu\gamma^\nu \psi \approx p^{\mu\nu}\ . 
\eeq
Namely, one tensor meson mode corresponds to the NG mode. 
In addition to the chiral $su_A(2)$-symmetry, the chiral $su_V(2)$-symmetry, namely $Q_V^1$ and $Q_V^2$, is also broken. 
Thus, from the second and the third equations in Eq.(\ref{6-12}), we obtain the other two NG modes as 
\beq\label{6-14}
& &{\bar \psi}\gamma^\mu\gamma^\nu \tau^{a=2}\psi \approx t^{\mu\nu}_2\ , \nonumber\\
& &{\bar \psi}\gamma^\mu\gamma^\nu \tau^{a=1}\psi \approx t^{\mu\nu}_1\ . 
\eeq
Namely, two tensor meson modes with $a=1$ and 2 are also the NG bosons. 
Therefore, under the tensor condensate, the chiral $su_V(2)$-symmetry is broken 
as well as the chiral $su_A(2)$-symmetry breaking similar to the case of the pseudovector condensate.
Thus, three NG modes are realized on the tensor condensate. 
This situations are exactly shown in section 5.2. 

If there exists the quark spin polarization under consideration in this paper, 
the chiral symmetry is broken by the condensation with respect to the third component of the isospin, $\tau_3$. 
In medium, the Lorentz symmetry is also broken by the quark spin polarization because the quark spin 
is polarized along $z$-axis, namely $\braket{\psi^{\dagger}\Sigma_{\mu=3}\tau_3\psi}$.
As for the Lorentz symmetry breaking, three NG modes should also appear. 
However, we impose a strong constraint, namely $\mu=3$ for spatial component and $i=3$ for 
the isospin component always pair. 
Thus, if the quark spin polarization, for example pseudovector condensate, rotates on planes including 
the third axis, $(i,3)=(0,3)$-, $(1,3)$- and $(2,3)$-plane in the isospin space, 
the pseudovector condensate simultaneously rotates on planes including the third ($z$-) axis, 
$(\mu,3)=(0,3)(=(t,z))$-, $(1,3)(=(x,z))$- and $(2,3)(=y,z))$-plane on the space-time.  
Therefore, three NG modes should disappear. 
When the generators for the symmetry transformation are linearly independent, 
the reducing the number of the NG modes occurs, which has been indicated in Ref.\cite{LM}.
However, it is interesting to know how many NG modes appear precisely in medium by the Lorentz symmetry breaking. \cite{WB} 
It is a future work to show which NG modes are type-I or II. 
This is an interesting problem because the behavior of the dispersion relation of the NG mode is determined in the 
long-wavelength limit.

\section*{Acknowledgements}

The authors would like to express their sincere thanks to Referee for his/her helpful and useful comments, 
especially about the number of the massless Goldstone modes and the appearance of the massless modes 
due to the meson mixing relating to the symmetry-breaking patterns. 
The authors also thank to Mr. K. Matsusaka and Mr. K. Yamashita, who 
were carried out the preliminary work with them about the massless modes on the pseudovector condensate 
and on the tensor condensate, respectively. 
The authors acknowledge 
to the member of the 
Many-Body Theory Group in Kochi University.



\appendix

\section{Detail calculations for the inverse propagators $\Gamma_{\rho,i\neq 3}^{33}(p\rightarrow 0)$, 
$\Gamma_{a,i\neq 3}^{33}(p\rightarrow 0)$. 
 }

By use of the anticommutation relation for the Dirac gamma matrices $\gamma^\mu$ and $\gamma_5$, 
and the isospin matrices $\tau_i$, Eq.(\ref{3-26}) or (\ref{3-27}) is transformed. 
Noting $(AB)^{-1}=B^{-1}A^{-1}$ for two matrices $A$ and $B$, we can calculate the inverse propagator : 
\beq\label{a-1}
& &{{\Gamma^{33}}_{\rho/a,i\neq 3}}(p\rightarrow 0)  \nonumber\\
&=&\frac{2}{G}-4{\int}\frac{d^4k}{i(2\pi)^4}\mathrm{tr}\left[({\gamma^3}{\tau_i})
\frac{1}{\fsl{k}-2{a_0}{\gamma_5}{\gamma^3}{\tau_3}}({\gamma^3}{\tau_i})
\frac{1}{\fsl{k}-2{a_0}{\gamma_5}{\gamma^3}{\tau_3}}\right] \nonumber\\
&=&\frac{2}{G}-4{\int}\frac{d^4k}{i(2\pi)^4}\mathrm{tr}
\left[({\gamma^3}{\tau_i})\frac{1}{(\fsl{k}+2{a_0}{\gamma_5}{\gamma^3}{\tau_3})
(\fsl{k}-2{a_0}{\gamma_5}{\gamma^3}{\tau_3})}(\fsl{k}+2{a_0}{\gamma_5}{\gamma^3}{\tau_3}) \right. \nonumber \\
&  &\qquad\qquad\qquad\ \ \ \ 
\left.\times ({\gamma^3}{\tau_i})(\fsl{k}+2{a_0}{\gamma_5}{\gamma^3}{\tau_3})
\frac{1}{(\fsl{k}-2{a_0}{\gamma_5}{\gamma^3}{\tau_3})
(\fsl{k}+2{a_0}{\gamma_5}{\gamma^3}{\tau_3})}\right] \nonumber\\
&=&\frac{2}{G}-4{\int}\frac{d^4k}{i(2\pi)^4}\mathrm{tr}
\left[({\gamma^3}{\tau_i})\frac{1}{k^2-4{a_0}^2-4{a_0}{k_3}{\gamma_5}{\tau_3}}
(\fsl{k}+2{a_0}{\gamma_5}{\gamma^3}{\tau_3}) \right. \nonumber\\
& &\qquad\qquad\qquad\qquad\qquad \left.\times ({\gamma^3}{\tau_i})(\fsl{k}+2{a_0}{\gamma_5}{\gamma^3}{\tau_3})
\frac{1}{k^2-4{a_0}^2+4{a_0}{k_3}{\gamma_5}{\tau_3}}\right] \nonumber\\
&=&\frac{2}{G}-4{\int}\frac{d^4k}{i(2\pi)^4}\mathrm{tr}
\left[({\gamma^3}{\tau_i})\frac{1}{(k^2-4{a_0}^2+4{a_0}{k_3}{\gamma_5}{\tau_3})
(k^2-4{a_0}^2-4{a_0}{k_3}{\gamma_5}{\tau_3})}\right. 
\nonumber\\
& &\left.\qquad\qquad\qquad
\times
(k^2-4{a_0}^2+4{a_0}{k_3}{\gamma_5}{\tau_3})(\fsl{k}+2{a_0}{\gamma_5}{\gamma^3}{\tau_3})\right. 
({\gamma^3}{\tau_i})(\fsl{k}+2{a_0}{\gamma_5}{\gamma^3}{\tau_3}) 
\nonumber\\
& &\qquad\qquad\qquad
\left.\times 
(k^2-4{a_0}^2-4{a_0}{k_3}{\gamma_5}{\tau_3})\frac{1}{(k^2-4{a_0}^2+4{a_0}{k_3}{\gamma_5}{\tau_3})
(k^2-4{a_0}^2-4{a_0}{k_3}{\gamma_5}{\tau_3})}\right]  \nonumber\\
&=&\frac{2}{G}-4{\int}\frac{d^4k}{i(2\pi)^4}\frac{1}{\left[(k^2-4a_0^2)^2-16{a_0}^2{k_3}^2\right]^2}  \nonumber\\
& &\qquad\qquad\qquad
\times \mathrm{tr}\left[{\gamma^3}{\tau_i}\left\{(k^2-4a_0^2){\fsl{k}}+8{a_0}^2{k_3}{\gamma^3}
\right. \right.
\left.
+[4{a_0}{k_3}{\gamma_5}
{\fsl{k}}+2{a_0}(k^2-4a_0^2){\gamma_5}{\gamma^3}]{\tau_3}\right\}{\gamma^3}{\tau_i}
\nonumber\\ 
& &\qquad\qquad\qquad\quad
\times\left\{(k^2-4a_0^2){\fsl{k}}+8{a_0}^2{k_3}{\gamma^3}
\right.
\left.\left.
+[-4{a_0}{k_3}{\fsl{k}}{\gamma_5}+2{a_0}
(k^2-4a_0^2){\gamma_5}{\gamma^3}]{\tau_3}\right\}\right]  . 
\eeq
In the second line of the last equality, moving $\gamma^3\tau_i$ by using the anticommutation relations, 
trace is calculated. 
Here, noting the following relations, namely 
\beq\label{a-2}
& &{\rm tr}\ \tau_i=0\ , \qquad {\rm tr}\ \tau_i^2=1\ , \nonumber\\
& &\mathrm{tr}(\gamma^3{\fsl{k}})={k_\mu}\mathrm{tr}({\gamma^3}{\gamma^\mu})=-4k_3 \nonumber\\
& &\mathrm{tr}(\gamma^3{\fsl{k}}\gamma^3{\fsl{k}})
={k_\mu}{k_\nu}\mathrm{tr}(\gamma^3{\gamma^\mu}\gamma^3{\gamma^\nu})
={k_\mu}{k_\nu}\mathrm{tr}\left[(2g^{\mu3}-{\gamma^\mu}{\gamma^3})(\gamma^3\gamma^\nu)\right]
\nonumber\\
& &\qquad\qquad\ \ \ 
=-2{k_3}{k_\nu}\mathrm{tr}(\gamma^3\gamma^\nu)+{k_\mu}{k_\nu}\mathrm{tr}({\gamma^\mu}{\gamma^\nu}) 
=8{k_3}^2+4k^2\ , \nonumber\\
& &{\rm tr}\ (\gamma^3\fsl{k}\gamma^3)={\rm tr}\ (\gamma^3\fsl{k}\gamma_5) =
{\rm tr}\ (\gamma_5\gamma^\mu)={\rm tr}\ \fsl{k}=0\ , 
\eeq
and after some tedious calculations for trace, Equstion (\ref{a-1}) can be simply expressed as  
\beq\label{a-3}
{{\Gamma^{33}}_{\rho/a,i\neq 3}}(p\rightarrow 0)
&=&\frac{2}{G}-96{\int}\frac{d^4k}{i(2\pi)^4}\frac{2{k_3}^2+k^2-4{a_0}^2}{\left[(k^2-4{a_0}^2)^2-16{a_0}^2{k_3}^2\right]}\ .
\eeq

\end{document}